# Coherent Phonons, Localization and Slow Polaron Formation in Lead-free Gold Perovskite

Sankaran Ramesh[1*], Yonghong Wang[2], Pavel Chabera[1], Rafael Araujo[3], Mustafa Aboulsaad[3], Tomas Edvinsson[3], Feng Gao[2], Tönu Pullerits[1*]

[1] Division of Chemical Physics and NanoLund, Lund University, Box 124, 221 00 Lund, Sweden

[2] Department of Physics, Chemistry, and Biology (IFM), Linköping University, Linköping, 581 83, Sweden

[3] Department of Materials Science and Engineering – Solid State Physics, Uppsala University, Box 534, SE-75121 Uppsala, Sweden

*Correspondence: sankaran.ramesh@chemphys.lu.se, tonu.pullerits@chemphys.lu.se

## Abstract

Lead-free metal halide perovskites are emerging as less-toxic alternatives to their lead-based counterparts. However, their applicability in optoelectronic devices is limited, and the charge transport dynamics remain poorly understood. Understanding photo-induced charge and structural dynamics is critical for unlocking the potential of these novel systems. In this work, we employ ultrafast optical and Raman spectroscopy combined with band structure calculations to investigate the coupled electronic and vibrational dynamics in Caesium gold bromide, a promising lead-free perovskite. We find that the band-edge charge transfer states are strongly coupled to Au-Br stretching phonon modes, leading to frequency modulation of absorption by impulsively excited coherent phonons. Early-stage relaxation is characterized by dynamics of delocalized charge transfer excitation and slowly decaying coherent phonons. The electronic and vibrational relaxation reveals a slow formation of a localized polaronic state in the 10-20 ps timescale. Using a displaced harmonic oscillator model, the polaronic binding energy is estimated to be ~80 meV following lattice relaxation along the phonon modes. Strong exciton-phonon coupling and slow polaron formation via coupling to lattice modes make this material a promising testbed for the control of coherent phonons and localized polaronic states using light.

## 1. Introduction

Metal halide perovskites are semiconductors with exceptional optical properties, making them excellent materials for solar energy harvesting and light emission.[1, 2] They are characterized by strong electron-phonon coupling, which gives rise to rich photo-induced dynamics.[3] Examples of such phenomena include the hot phonon bottleneck effect,[4] where the electronic relaxation is modified by equilibrium with hot phonons; the formation of phonon-dressed electronic states such as self-trapped carriers or polarons;[5-8] and photo-induced structural phase transitions.[9] These are at the core of optoelectronic technologies such as hot carrier solar cells, broad-band white light emitters and ultrafast optical-switching.[10-12] The electron-phonon coupling could provide a means to manipulate material functionality through light and lattice modes.

Photoexcitations and the related charge carriers in lead-based perovskites are widely studied and have highly efficient transport properties.[13-15] However, challenges of environmental toxicity and stability have spurred research into alternative compositions. This includes a variety of lead-free systems where Sn, Ge, Ag, Bi, Sb, etc, replace the Pb.[16-18] Several lead-free candidates also exhibit strong electron-phonon interaction that influences their excited state relaxation. The ultrafast formation of a localized polaronic state, where the lattice around the charge relaxes into a locally distorted configuration, has been reported in many cases. The localized state has lower mobility, which could hamper its wide applicability.[18-20] This presents a challenge for their use in photovoltaic devices compared to conventional Pb-based perovskites that generally have better efficiencies. However, opportunities for its use in other forms, including thin-layered light absorbers, thermochromic devices and white-light emitters have been highlighted in previous studies.[21-24]

The localization process in this broad class of materials remains poorly studied, with scarce spectroscopic investigations. Yet, a clear understanding of ultrafast electronic and structural dynamics is required to identify and tune novel lead-free perovskites and perovskite-inspired materials for practical applications. In double perovskites with the general formula $A_2MM'X_6$, it has been proposed that localization occurs due to low electronic dimensionality,[25] which denotes the low spatial overlap between orbitals of its conduction band minimum and valence band maximum. The charge localization time in $Cs_2AgBiBr_6$ double perovskite was estimated to be 1 ps by Wright et al.[19] and 5 ps by Wu et al.[20] using time-resolved optical and terahertz spectroscopies. Studies by Yang et al.[26] in $Cs_2AgSbBr_6$ and by Buizza et al.[27] in in $Cu_2AgBiI_6$ using ultrafast methods have also showed the formation of the polaronic state in 1-2 ps after photoexcitation. As localization involves the coupled relaxation of charge and lattice, it necessitates the use of multiple complementary techniques to map out the dynamics.

Given the challenges associated with lead-free double perovskites, significant research has focused on developing these materials. Inorganic Caesium gold halide perovskites ($Cs_2Au_2X_6$ where X is a halide ion) are one such class of lead-free perovskites predicted to make good photovoltaic materials due to strong absorption with a bandgap in the near-IR region and weakly bound excitons.[28] In its ground state, the Au cations in $Cs_2Au_2Br_6$ have mixed oxidation states causing distortion of the charge distribution on the Au ions and forming a double-perovskite crystal structure. This results from the interplay between Coulomb interactions and the Jahn-Teller effect.[29] Photoinduced and pressure-induced phase transitions have been observed due to the structure's properties depending on the charge ordering of the Au ions.[30,31] It has also been reported that the structural transitions considerably affect its conductivity.[32] Nonetheless, the interactions between charge carriers and the lattice, as well as the localization phenomena in these systems, are not yet well comprehended. Investigating these ultrafast dynamics using advanced spectroscopic techniques can provide valuable insights into the mechanisms governing charge mobility and stability. Such knowledge is crucial for tailoring these materials for specific applications, enhancing their efficiency, and expanding their potential uses in next-generation optoelectronic devices.

In this work, we investigate the non-equilibrium electronic and vibrational dynamics in caesium gold bromide thin films using ultrafast and Raman spectroscopy. Pump-probe measurements show that, at early times after photoexcitation, a derivative-shaped signal is formed due to band-edge renormalization. Pronounced spectral oscillations appear due to modulation of the absorption by impulsively excited coherent phonons. Density functional theory and resonance Raman spectroscopy are utilized to clarify the electronic and vibrational states. Together with transient absorption, these techniques reveal that the characteristic inter-valence charge transfer (CT) states at the band-edge of $Cs_2Au_2Br_6$ are coupled to its Au-Br stretching LO phonon modes via a large displacement of the potential energy surfaces. By analyzing the electronic and vibrational dynamics, we track the slow formation of a localized polaronic state in the 10-20 ps timescale, with a lattice relaxation energy of ~80 meV. These findings show that electron-phonon coupling could be leveraged to control charge carrier localization and coherent lattice dynamics in this emerging class of materials.

## 2. Results

### 2.1 Optical Properties

Polycrystalline Caesium gold bromide films were synthesised by spin-coating, as described in the *Methods* section. In this material, the Jahn-Teller effect causes the two neighbouring $Au^{2+}$ to undergo disproportionation by re-arranging charges forming the more stable $Au^+$ and $Au^{3+}$.[33] The material therefore has the chemical formula $Cs_2Au^{3+}Au^+Br_6$ or $Cs_2Au_2Br_6$ and a distorted octahedral perovskite structure (schematically shown in Figure 1a). The distorted crystal comprises compressed $[Au^{3+}Br_4]^-$ octahedra and elongated $[Au^+Br_2]^-$ octahedra.

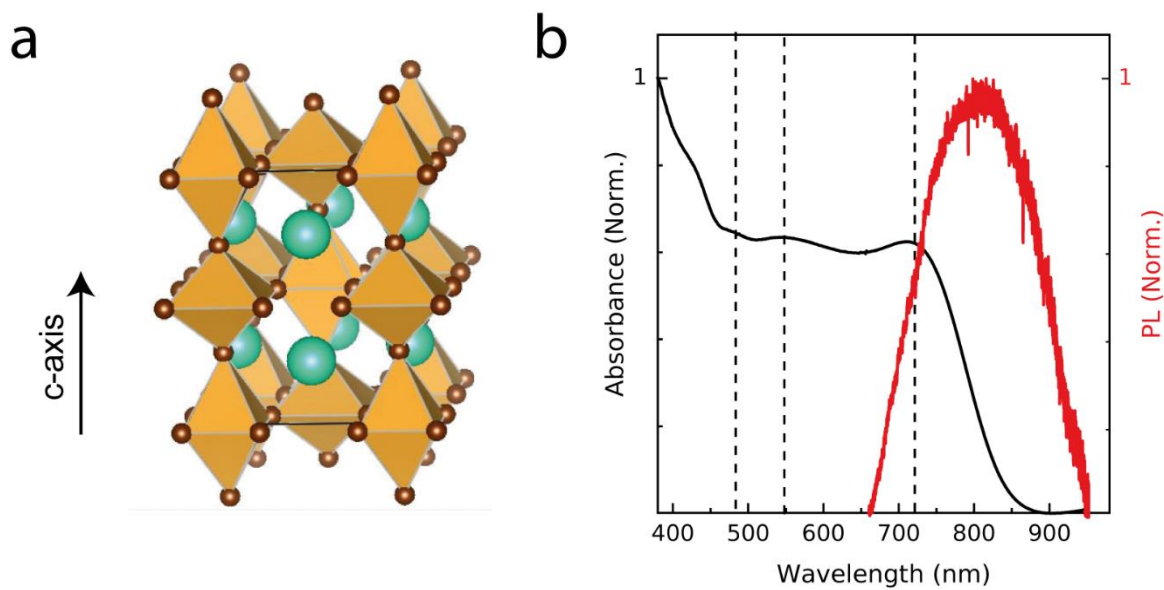

*Figure 1: Structure and absorption spectrum of $Cs_2Au_2Br_6$ thin films: (a) Schematic crystal structure. The blue spheres are Cs, The brown spheres are Br and the Au ions are at the center of the octahedra (not visible in the picture) (b) Absorption and photoluminescence spectrum (excitation wavelength 500 nm). The dashed lines indicate the inter-valence charge transfer transitions.*

The absorption spectrum of $Cs_2Au_2Br_6$ (Figure 1b) show light absorption across the whole visible spectrum. The electronic structure of $Cs_2Au_2Br_6$ was computed by density functional theory (DFT) using Projected Augmented Waves (PAWs). The calculations confirm the formation of mixed oxidation states of Au and distorted octahedral structure as expected. (details of the calculations are given in Supplementary Notes 1-2). The band structure shows that $Cs_2Au_2Br_6$ has an indirect band gap of 1.32 eV, with a direct gap close in energy (1.35 eV) at the N-point of the Brillouin zone. This is well-aligned with previous studies[34, 35] and the computed band gap is close to the experimental value of 1.40 eV obtained here (Figure S1). The material has weak emission with a broad spectrum. (Figure 1b).

The absorption spectrum has features at 720 nm (~1.73 eV), 550 nm (~2.25 eV) and 490 nm (~2.53 eV), marked with dashed lines in Figure 1b. These are assigned to inter-valence transitions with strong charge transfer (CT) character between Au$^+$ 5d$_{x^2-y^2}$ to Au$^{3+}$ 5d$_{x^2-y^2}$ orbitals (720 nm), Au$^+$ 5d$_{yz,zx}$ to Au$^{3+}$ 5d$_{x^2-y^2}$ orbitals (550 nm) and Au$^+$ 5d$_{xy}$ to Au$^{3+}$ 5d$_{x^2-y^2}$ orbitals (490 nm) based on previous reports.[29, 36] In single crystals of Cs$_2$Au$_2$Br$_6$, it was reported that the optical response is dominated by the CT transitions between Au$^+$ and Au$^{3+}$ in the ab-plane. SEM images of the polycrystalline thin films (Figure S2) indicate the formation of microcrystalline grains with random orientations and sizes in the order of 100 nm. Even here the optical response is dominated by the CT transitions along the ab-plane.

## 2.2 Transient Dynamics at the Band-edge

We studied the excited state relaxation dynamics in the films using transient absorption (TA) spectroscopy. The films were excited by 70 fs pulses with a centre wavelength of 500 nm, and a delayed probe pulse recorded the transient differential absorption. Details of the experimental setup are given in the *Methods* Section. Figure 2a shows the pseudo-colour TA spectrum of Cs$_2$Au$_2$Br$_6$. Key features of the spectrum include negative amplitude ground-state bleach (GSB) signals at positions of the inter-valence CT transitions. A third negative feature at 430 nm was previously assigned to a ligand-to-metal charge transfer transition from the halide ion to the Au$^{3+}$ ion.[29] Excited-state absorption (ESA) signals with positive amplitude are also observed at 840 nm, 660 nm and 470 nm. Figure 2b provides slices of the TA spectrum at chosen delay times 0.3 ps, 1 ps, 20 ps and 500 ps, showing the electronic relaxation. Close to the band edge (~730 nm), at early time there is a derivative-shaped feature characterized by adjacent negative and positive peaks. The shape of the feature remains largely unchanged during its decay of about 20 ps. At the later time of 500 ps, the positive peak has decayed completely, and a red-shifted negative peak remains. A detailed discussion of the electronic dynamics at the band edge is given later, following the discussion of vibrational dynamics.

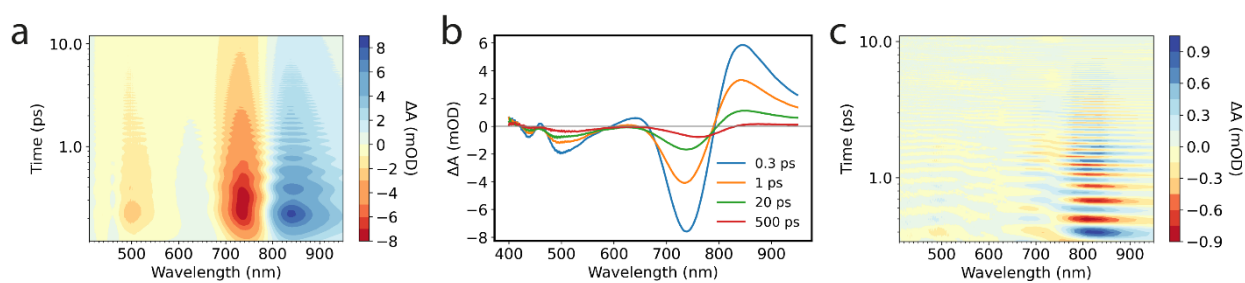

Figure 2: Transient absorption spectrum and vibrational properties of Cs$_2$Au$_2$Br$_6$ films. (a) pseudo-colour TA spectrum of Cs$_2$Au$_2$Br$_6$. (b) Slices of the TA spectrum at 0.3 ps, 1 ps and 20 ps. (c) Residual of the TA spectrum obtained after subtracting the electronic dynamics.

The spectrum in Figure 2a also shows a periodic beating pattern over the GSB and ESA bands at the band edge. To examine the beatings in detail, the non-oscillatory decay part of the TA spectrum is subtracted from the overall spectrum, and the residual oscillatory signal is obtained, as shown in Figure 2c. Pronounced spectral oscillations are observed close to the band edge, and the oscillation period remains unchanged over the probe wavelength range. There is a change in the phase of the oscillations at the centre of the GSB band. The amplitude of oscillations is the strongest at 830 nm where the ratio of oscillation amplitude $A_{osc}$ to the overall differential absorption signal $\Delta A$ is ~ 0.12. The pronounced amplitude of the oscillations with sub-ps periodicity and their decay in time suggests the excitation of high-amplitude coherent phonon modes which modulates the probe light absorption.[37-39]

2.3 Vibrational Properties

We investigated the vibrational properties of $Cs_2Au_2Br_6$ using Density Functional Perturbation Theory (Supplementary Note 3) and frequency-domain Raman spectroscopy. The phonon band structure and the density of states (DOS) was calculated (Figure S3). The results showed that the highest DOS of phonon modes is found between 10-50 cm$^{-1}$, while modes with a lower DOS are found at higher frequencies up to 200 cm$^{-1}$. The computed Raman spectrum of $Cs_2Au_2Br_6$ (Figure 3a) shows four Raman-active modes at 191 cm$^{-1}$, 163 cm$^{-1}$, 136 cm$^{-1}$ and 89 cm$^{-1}$, the latter with a very weak Raman cross-section. The highest frequency mode, 191 cm$^{-1}$, is associated with stretching Au-Br bonds along the c-axis in the elongated $[Au^+Br_2]^-$ octahedra. In contrast, the other three modes, including the one with the highest intensity (164 cm$^{-1}$), are linked to the stretching of Au-Br bonds along the ab-plane in the compressed $[Au^{3+}Br_4]^-$ octahedra. The schematic of these phonon modes in the crystal is shown in the inset and at the bottom of Figure 3a. Resonance Raman spectroscopy experiments showed three peaks, see Figure 3b. The frequencies of the modes in the experiment, 174 cm$^{-1}$, 183 cm$^{-1}$ and 207 cm$^{-1}$, are in good correspondence to the high-frequency modes from the theory. Hence these are assigned to different longitudinal optical (LO) phonon modes: the $[Au^{3+}Br_4]^-$ $B_{1g}$ mode, $[Au^{3+}Br_4]^-$ $A_{1g}$ mode and the $[Au^+Br_2]^-$ $A_{1g}$ mode, respectively.[32] The fourth and lowest frequency Raman-active mode from calculations, at 89 cm$^{-1}$, corresponds to Au-Br stretching modes towards the $Cs^+$ ion, which is not observed in the experiment, possibly because the Raman susceptibility of the mode is too low.

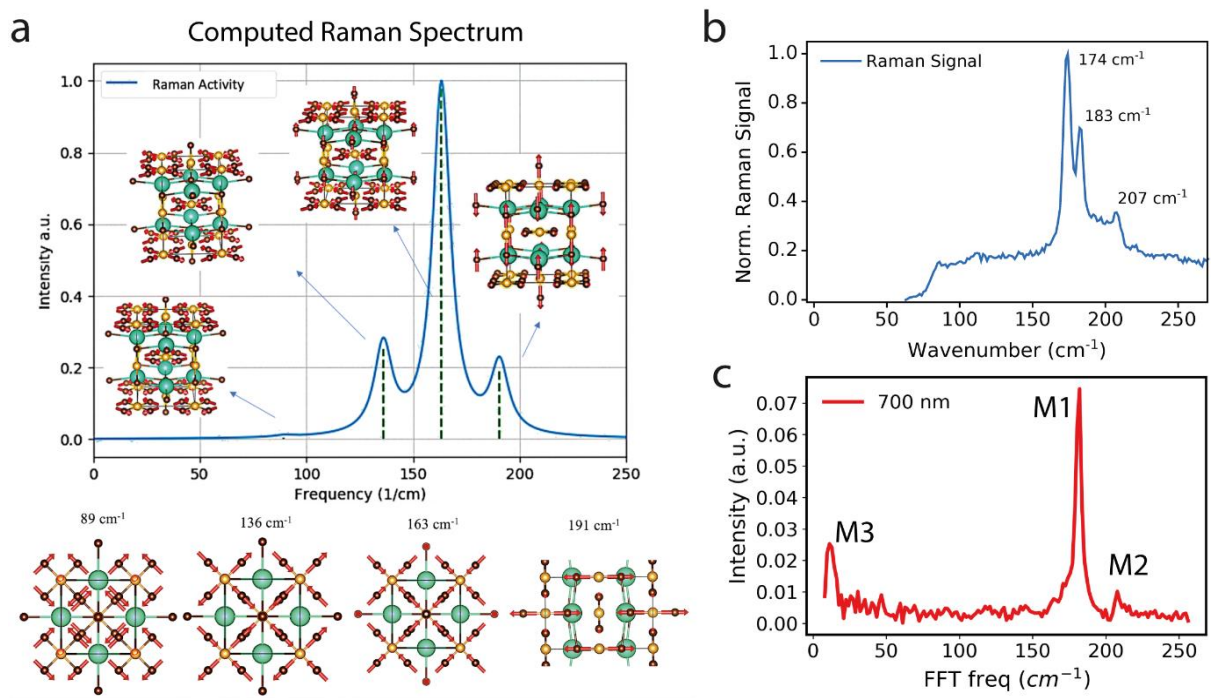

Figure 3: Vibrational properties of Cs$_2$Au$_2$Br$_6$. (a) The computed Raman spectrum of Cs$_2$Au$_2$Br$_6$ films showing four Raman-active modes and their associated nuclear motions in the crystal (side-view, top inset and top-view showing the ab- crystal plane, bottom) (b) Resonant Raman spectrum of Cs$_2$Au$_2$Br$_6$ with an excitation wavelength of 532 nm. (c) Fourier transformed residual TA spectrum of Cs$_2$Au$_2$Br$_6$ at 700 nm. Modes M1 and M2 are LO phonon modes related to the Au-Br stretching. M3 is an acoustic mode.

The Fourier transform of the residual TA spectrum (Figure 2c) along the delay times from 0.3 ps to 20 ps is shown as a 2D spectrogram in Figure S4 and a slice of the spectrogram at 700 nm probe wavelength is plotted in Figure 3c. The peaks are at frequencies ~182 cm$^{-1}$, ~207 cm$^{-1}$ and ~13 cm$^{-1}$ (denoted as M1, M2 and M3, respectively). The most intense peak at 182 cm$^{-1}$ (M1) and a weaker peak at 207 cm$^{-1}$ (M2) correspond to the Raman active A$_{1g}$ modes of the [Au$^{3+}$Br$_4$]$^-$ and [Au$^+$Br$_2$]$^-$ octahedra identified earlier. A weak shoulder feature on the lower frequency side of the M1 (182 cm$^{-1}$) peak indicates that the [Au$^{3+}$Br$_4$]$^-$ B$_{1g}$ mode overlaps with the more intense peak. The mode M3 at 13 cm$^{-1}$ is assigned to a coherent acoustic phonon, which is outside the observation window of the frequency-domain Raman measurement (Figure 3b). The coherent interaction of a short light pulse with Raman-active vibrational modes in a crystal can lead to the impulsive excitation of nuclear wavepackets.[40] The assignment of oscillations in the TA spectrum to coherent oscillations of M1 and M2 vibrational modes is confirmed by the good agreement between the theoretical calculations and time- and frequency-domain experiments.

The coherent phonons persist for several ps, as seen from the dynamics of the residual TA signal in Figure 4a. The residual oscillations are fitted using the equation:

$$y(t) = \sum_i A_{osc}^i \left[\cos(\Omega_i t + \theta_i)\right] \exp\left(-\frac{t}{\tau_i}\right), \quad (1)$$

where $y(t)$ is the intensity of the residual signal, $a_n$ is the amplitude of the $n^{th}$ oscillatory component of the residual signal and $\Omega_i, \tau_i, \theta_i$ are the frequency, decay time and phase of the $n^{th}$ oscillating component, respectively. The residual oscillations at 700 nm are modelled with $i = 3$, since we identified three coherent phonon modes in Figure 3c. The beating signal and the model fit are shown in Figure 4a. Figure 4b shows the individual frequency components of the fit, corresponding to the phonon modes. The optimized fit parameters are given in Table 1.

*Table 1 Parameters of Decay dynamics of coherent phonons*

|  | M1 | M2 | M3 |
|---|---|---|---|
|  | 182 cm$^{-1}$ | 207 cm$^{-1}$ | 13 cm$^{-1}$ |
| Amplitude $A_{osc}$ (mOD) | 0.240 ± 0.004 | 0.075 ± 0.006 | 0.010 ± 0.004 |
| Dephasing Time τ (ps) | 5.7 ± 0.2 | 2.6 ± 0.3 | 8 ± 5 |

The initial phase of the oscillations, represented by $\theta_i$ is (π/2). The fitted phase undergoes a shift of π near the GSB wavelength, as was expected. The slow dephasing of M1 and M2 modes can be rationalized by considering the phonon band structure and the vibrational density of states (DOS) versus the mode frequency in Cs$_2$Au$_2$Br$_6$ (Figure 4c). Purple and green colour lines in Figure 4c show the coherent phonon frequencies $\omega_{M1}$ and $\omega_{M2}$. The most significant contribution to the anharmonic decay of phonons is via the scattering of LO phonons with two acoustic phonons through an energy and momentum-conserving process called the Klemens decay.[41] In this three-phonon scattering process, one LO phonon decays into two LA phonons of equal energy. From the calculations, the phonon DOS is higher in the low frequency region (< 50 cm$^{-1}$), and the modes at half-frequency ($\omega_{M1}/2$) and ($\omega_{M2}/2$) lie in a region with low DOS (> 50 cm$^{-1}$). This implies that in the presence of a large population of non-equilibrium phonons of frequency $\omega_{M1}$ and $\omega_{M2}$, as in our case, the Klemens process is hindered due to low availability of the modes at half the frequency. Therefore, the lattice relaxation must proceed through higher-order processes involving four or more phonons, that have low probability. We point out that this explanation cannot rationalize the very long decay rate of coherent low frequency M3 phonons since at that frequency region density of phonon states is high. Coherent acoustic phonons with very long dephasing times have been observed before in perovskite materials.[42]

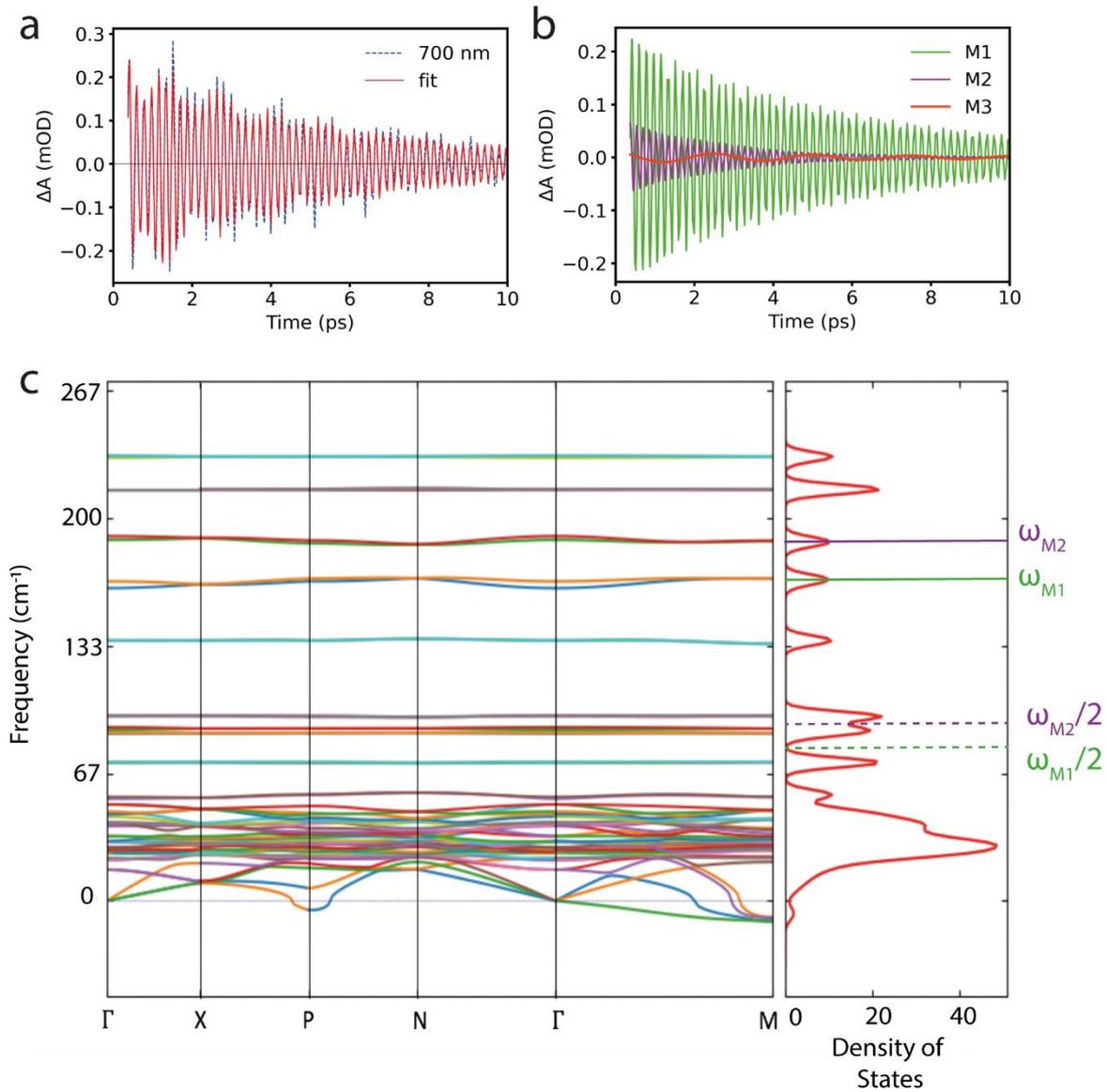

Figure 4: Decay of coherent phonons (a) Residual TA signal decay at 700 nm and fit using equation 1. (b) Contribution of each coherent phonon mode to the residual oscillations. (c) Calculated Phonon bandstructure of $Cs_2Au_2Br_6$ (Left) with density of states (Right). Purple and Green lines mark the frequencies corresponding to M1 ($\omega_{M1}$) and M2 ($\omega_{M2}$) modes, respectively. Dashed lines indicate half-frequency of the modes ($\omega_{M1}/2$ and $\omega_{M2}/2$).

## 2.4 Modulation of Electronic Properties by Coherent Phonons

We will now elucidate the electron-phonon interactions in the material from the TA dynamics. First, we identify the excitation mechanism of the coherent phonons. Impulsive stimulated Raman scattering is a process by which the crystal interacts with a short, broadband optical pulse.[40] It involves coherent interaction of the system with two components of the pulse separated in frequency by the phonon energy $\hbar\Omega$. The coherent phonons in turn modulate the electronic energy. The stimulated Raman scattering of the incident light pulse by the coupled electron-lattice system and the resulting phonon amplitude Q can be described using the driven oscillator equation:

$$\frac{d^2Q}{dt^2} + 2\Gamma\frac{dQ}{dt} + \Omega Q = F/\mu, \quad (2)$$

where $\Gamma$ and $\mu$ are the damping constant and reduced mass, respectively. F is the driving force for the oscillations, which can be written in terms of the dielectric constant of the medium. As the measure of electron-phonon coupling, we consider the deformation potential constant $\Xi$, which is the change in electronic energy with the ionic displacement due to the phonons. The phonon mode amplitude Q(t) is obtained from the solution of equation ( 2 ) as (details in Supplementary Note 4)[37, 43, 44]

$$Q(t) = \frac{C\varepsilon_{Im}}{\Omega_1^2} \Xi\, I_0\, e^{-\Gamma t}\cos(\Omega_1 t + \theta), \quad (3)$$

where C is a constant pre-factor, $\varepsilon_{Im}$ is the imaginary part of the dielectric constant, $\Omega_1 = \sqrt{\Omega^2 - \Gamma^2}$, $\theta = \tan^{-1}\left(\frac{d\varepsilon_{Re}}{d\omega}\Big/\frac{2\varepsilon_{Im}}{\Omega_1}\right)$, $I_0 = \int_{-\infty}^{\infty}|E(t)|^2 dt$ is the integrated pulse intensity and $\Gamma$ is the dephasing rate. The differential absorption signal from the modulation is proportional to the mode amplitude Q(t). The residual TA signal $\Delta A_{osc}$ is then obtained as:

$$\Delta A_{osc} \propto \frac{d\varepsilon_{Im}}{dE} \frac{\Xi^2}{\Omega_1^2} I_0 . \quad (4)$$

$\Delta A_{osc}$ has a linear dependence on $\frac{d\varepsilon_{Im}}{dE}$ as well as the incident pulse intensity $I_0$. Figure 5a shows the plot of the residual TA spectrum at an arbitrarily chosen probe delay of 1.2 ps. For the consistency check, we present in the same figure $\frac{d\varepsilon_{Im}}{dE}$ obtained from the absorption spectrum using the Kramers-Kronig relations (Supplementary Note 5). The shape of the plots is very similar, confirming that the $\Delta A_{osc}$ behaves as predicted by Eq. (4). Further, Figure 5b shows $\Delta A_{osc}$ at a fixed probe delay time of 3.5 ps as a function of laser pulse intensity. The trend is linear, confirming the expected behaviour with pump intensity. These observations, along with the agreement with the resonance Raman experiments, confirm that the coherent phonons in Cs$_2$Au$_2$Br$_6$ are excited by the resonant impulsive stimulated Raman process and result in the oscillating residual TA signal in our experiment.

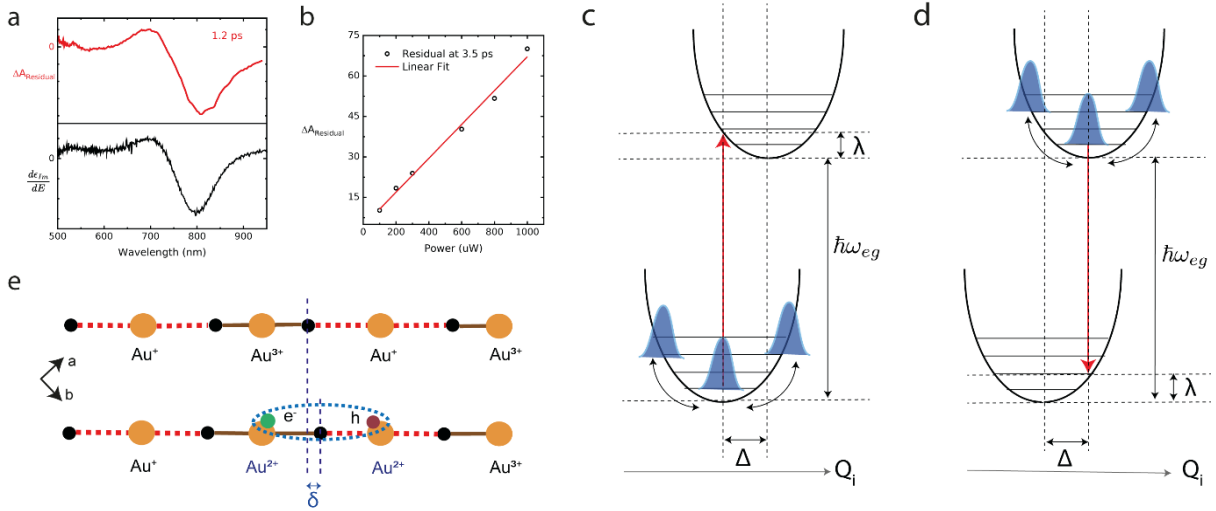

*Figure 5: Coherent LO Phonons in Cs$_2$Au$_2$Br$_6$ : (a) Top: Residual TA spectrum at the delay time of 1.2 ps. Bottom: derivative of the imaginary part of the dielectric constant, obtained from the steady-state absorption spectrum using Kramers-Kronig relations. (b) The residual TA signal at 3.5 ps delay time, measured for different pump beam intensities, in µW. (c-d) Illustration of formation of a vibrational coherence on the ground state potential energy surface (PES) (c) and on the excited-state PES (d). Δ is the displacement between the PESs and λ is the re-organization energy. Q$_i$ is a generalized nuclear co-ordinate and E$_g$ is the band-gap. (e) Schematic of the crystal structure in the ab-plane. Top row shows the nuclear arrangement in the ground state, with the black and dashed red arrows indicating the different Au-Br bond lengths. Bottom row shows the formation of a CT state (in the blue dashed lines) and the nuclear motion about a new equilibrium position δ.*

Considering the scenario of a single charge interacting with lattice vibration, a displaced harmonic oscillator picture can be used to understand the modulation of electronic energy by the coherent phonons.[45] This is shown in Figure 5c-d. The ground and excited state corresponding to the transition are represented as harmonic potentials with displaced equilibrium positions (Δ) along the nuclear coordinate corresponding to the phonon mode $M_i$ ($i = 1,2,3$). $\hbar\omega_{eg}$ is the electronic band gap and $\lambda_i$ is the lattice reorganization energy. Laser pulse excitation leads to the formation of a superposition of vibrational states or a wavepacket. The vibrational wavepacket can be formed on the ground state potential energy surface (PES) (Figure 5c) or the excited state (PES) (Figure 5d), depending on the type of interaction with the pump photon. As the wavepacket evolves on a PES, its interaction with the probe pulse leads to periodic modulation of the differential absorption. The modulation is most intense at the wavepacket's turning points. In contrast, the modulation amplitude is small approaching zero at around the potential surface minimum.[46] In the two possible scenarios shown schematically, the zero of the modulation amplitude could be centred either at the GSB (Figure 5c) or at a stimulated emission band (Figure 5d). In Figure S5, we analyze the Fourier-transformed residual TA spectrum at M1, M2 and M3 frequencies. The zero is at the GSB energy for M1 and M2 modes, indicating that the vibrational coherence propagates on the ground state PES. The center of modulation by M3 is red-shifted, indicating that the coherent phonons of mode M3 propagate along the excited-state PES and therefore modulate the stimulated emission signal.

This analysis can be complemented with a physical understanding of the strong electron-phonon coupling. The nuclear displacements of the LO phonon modes M1 and M2 are associated with the stretching motion of the Au-Br bonds, which lie along the same direction as the $Au^+$ to $Au^{3+}$ CT state excited by light. Figure 5e shows a schematic of the crystal structure in the ab-plane. The blue dotted circle indicates a CT state and the motion of Au-Br stretching modes is shown by the nuclear displacement δ. The alignment of the electric fields from the electronic transition and the nuclear motion along the same axis rationalizes the strong electron-phonon coupling in $Cs_2Au_2Br_6$.

## 3. Discussion

A combined picture of the electronic and vibrational dynamics in $Cs_2Au_2Br_6$ can be presented based on the results. Photoexcitation generates a state with strong charge transfer character owing to the electron population redistribution between orbitals of Au in $[Au^{3+}Br_4]^-$ and $[Au^+Br_2]^-$ octahedra. We envision that the state is initially delocalized over several units. Coherent vibrational wave packets related to the Au-Br bond stretching motion are also generated due to a resonant impulsive stimulated Raman process. Considering the pronounced coherent phonon modulation of the band edge TA signal, we propose that the CT state interacts with the lattice vibrations along these modes to relax to a localized polaronic state.

The polaron binding energy is estimated by employing a widely-used model based on displaced harmonic oscillator potentials that compares the amplitude of oscillations in the residual spectra ($A_{osc}^i$) to the derivative of the sample absorbance.[47-50] (We discuss how this model is obtained in the context of a localized charge in Supplementary Note 4) The Huang-Rhys factor ($S_i$) is a dimensionless parameter that represents the strength of the electron-phonon coupling of a mode $M_i$. The displacement of the two surfaces ($\Delta_i$), the lattice reorganization energy ($\lambda_i$) and the Huang-Rhys factor ($S_i$) are determined using the expression

$$\lambda_i = \frac{A^i_{osc}}{\left(\frac{dA_0}{dE}\right)}, \quad (5)$$

where $A_0$ is the sample absorption spectrum normalized to the intensity of the GSB at zero-time delay. $A^i_{osc}$ is obtained for each mode from the equation ( 1 ). The parameters $S_i$, $\Delta_i$ and $\lambda_i$ are related to each other: $S_i = \Delta_i^2/2$ and $\lambda_i = S_i \hbar \Omega_i$ ($\Omega_i$ = frequency of the phonon mode). The obtained parameters for the LO phonon modes M1 and M2 are shown in Table 2.

Table 2: Exciton-Phonon Coupling Parameters

| Mode ($M_i$) | $S_i$ | $\Delta_i$ | $\lambda_i$ (meV) |
|---|---|---|---|
| M1 (182 cm$^{-1}$) | 1.40 | 1.67 | 31.6 |
| M2 (207 cm$^{-1}$) | 0.27 | 0.74 | 7.0 |

The Huang-Rhys factors indicate strong electron-phonon coupling and large displacement of PESs compared to other metal halide perovskites[48], but closer to the values found in perovskites with 2D and quasi-1D crystal structures.[50] The total reorganization energy due to lattice relaxation along the M1 and M2 reaction coordinates is $\lambda_{LR} = 2(\lambda_{M1} + \lambda_{M2}) \approx 78$ meV. Hence, we infer that the lattice reorganization energy directly gives the polaron binding energy. The lattice relaxation and formation of the localized polaronic state proceeds along the nuclear coordinates of the M1 and M2 modes $Q_{M1}$ and $Q_{M2}$. Using the same model for the acoustic mode M3 we obtained a Huang-Rhys factor of 1.03, $\Delta$ of 1.44 and a reorganization energy $\lambda$ of 1.7 meV, which represents a small additional contribution to the overall lattice relaxation.

To elucidate the details of the photoexcitation dynamics, we analyze the evolution of the TA spectrum. Figure 6a shows the TA spectrogram around the band-edge separated into the early stage up to 20 ps and the later stage. As discussed in Section 2.2, the GSB and ESA bands form a derivative-shaped feature in the early stage. Only a red-shifted negative feature remains at later delay times, as shown by the black arrow (Figure 6a). Figure 6b shows the kinetics of the GSB and ESA, at 755 nm and 845 nm, respectively (sign of the GSB is inverted and time-zero has been shifted to ~0.2 ps to facilitate the logarithmic scale representation). In the initial stage, the GSB and ESA dynamics follow each other (Figure 6b). We assign the derivative-shaped feature to renormalization of the band-edge electronic energy by the local field of the photoexcited CT states, similar to that reported in other semiconductors including metal halide perovskites.[49-52] Decay of the derivative feature suggests smearing out of the charge transfer character of the excited state, and a long-lived component remains at 755 nm.

We model the derivative-shaped feature in time using two Gaussian peaks (Supplementary Note S6) revealing the red-shift of the centre position of the negative peak.

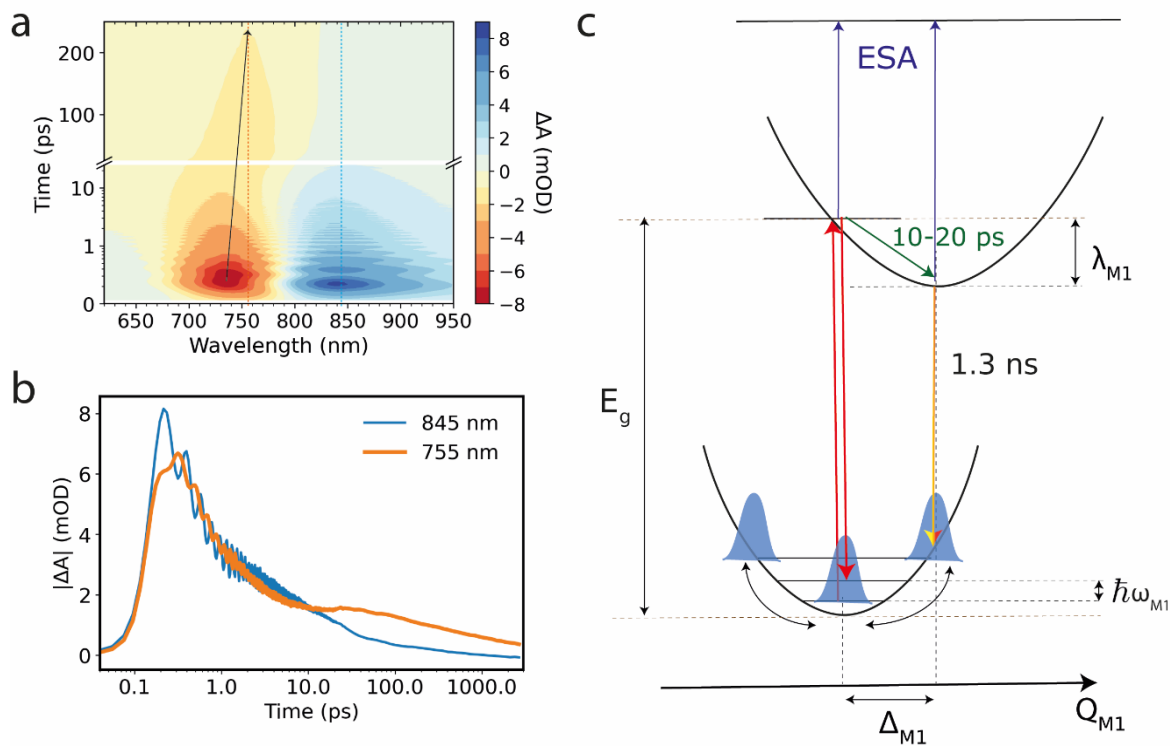

Figure 6: Dynamics of the band-edge CT state (a) pseudo-color TA spectrum near the band-edge showing the red-shift of the GSB peak (black arrow). (b) Kinetics near the band-edge GSB and ESA energies (wavelengths shown with dashed lines in panel (a)), showing the correspondence between the decays at early stage and the long-lifetime component in the GSB at later stage. (c) Schematic of the localization shown using the displaced harmonic oscillator picture along the nuclear co-ordinate $Q_{M1}$ of mode M1. The electronic excitation is generated in the conduction band (not shown), and it relaxes to the localized state over 10-20 ps. In the schematic, the top parabola is the potential energy surface of the polaronic state and the bottom parabola is the potential energy surface of the ground state. $E_g$ is the band-gap, $\lambda_{M1}$ re-organization energy and $\Delta_{M1}$ is the displacement along $Q_{M1}$.

To elucidate the dynamics leading to polaron formation, we performed global analysis of the TA data using singular value decomposition (Figure S9). The analysis yielded four decay components with lifetimes of 0.3 ps, 1.7 ps, 23 ps and 1.3 ns which are described in detail in Supplementary Note 7. The fastest 0.3 ps component is assigned to the thermalization of charges to form the initially delocalized excitation. We assign the 1.7 ps process to the decay of the delocalized CT state to a more localized state with less pronounced CT character leading to a decay of the derivative-shaped signal. Based on the damping time of coherent phonons we can set the timescale of the vibrational relaxation. Dephasing of the coherence is the result of the population decay and pure dephasing. Provided the Huang-Rhys factor of 1.4, the vibrational coherences mainly correspond to the density matrix elements $\rho_{01}$ and $\rho_{02}$. As the first approximation the 0-level has long lifetime which sets the lower limit (negligible pure dephasing) for the vibrational relaxation time as twice the dephasing time.[53] Thereby the coherent M1 phonon dephasing time of ~ 6 ps (*Table 1*) means that the vibrational relaxation of excitation along the M1 mode co-ordinate takes 12 ps or longer. This implies that the polaron formation would occur on the timescale of 10-20 ps. This is consistent with the ~20 ps component obtained from the global analysis. Therefore, the 20 ps process is assigned to the lattice relaxation and formation of the polaronic state. The propagation of coherent vibrational wave packets along the ground state PES and lattice relaxation is schematically shown in Figure 6c. To rationalize the red shift of the GSB and disappearance of the ESA in this timescale, we consider that an absorption of probe to a higher excited state (dark blue arrows in Figure 6c) contributes as a positives signal in TA. Polaron formation will then shift the ESA signal from its initial position to higher energy, causing overlap with the GSB and the apparent redshift of the negative peak.

We point out that the envisioned electronic delocalization has consequences to the experimental Huang-Rhys factors. Delocalization leads to lower values of effective measured electron-phonon interaction compared to the corresponding coupling in a localized state.[54] This means that the polaron reorganization energy of about 80 meV based on the initial excitation conditions should be taken as the lower estimate. The true reorganization energy of the localized polaron may be larger than that.

Following the decay of the coherent phonons, local crystal distortions remain, stabilising the polaronic state. The slowest 1.3 ns component from the global analysis is assigned to the recombination of the polaron back to the ground state where the lattice also returns to its original structure.

The localization process in other double perovskite compositions has also been described as due to interactions with optical and acoustic phonons. With the more commonly studied Bi-based double perovskites, the reported Huang-Rhys factors in these systems are generally an order of magnitude higher (S ~ 12 in $Cs_2AgBiBr_6$[55]). Spectroscopy measurements have described the self-trapping of charges in this system in the 1-5 ps timescale by strong coupling to optical and acoustic phonons.[19, 20] $Cs_2AgBiBr_6$ and other double perovskites with large Huang-Rhys factors are also characterized by large Stokes shifts and broad emission spectra.[8, 22, 56] In $Cs_2Au_2Br_6$, we find that the emission spectrum is weaker and has lower Stokes shift in comparison (Figure 1b), an observation consistent with the lower estimated Huang-Rhys factors. The emerging picture of localization process in this study presents an intermediate electron-phonon coupling accompanied by slow decay of coherent longitudinal and acoustic phonons.

In lead-based halide perovskites, the suppression of the Klemens decay channel has been proposed as an explanation for the hot phonon bottleneck effect,[4] where LO phonons generated by electron-phonon coupling experience lower rate of decay. This can lead to slow cooling of in-band charges to the band-edge. Gaining a clear picture of vibrational relaxation of the coherent phonons may also help in understanding the ultrafast dynamics of the hot phonon bottleneck effect and designing energy harvesting systems that minimize energy dissipation to the lattice.

## 4. Conclusion

In this study, we investigated the coupled electronic and vibrational dynamics in $Cs_2Au_2Br_6$ using ultrafast optical and Raman spectroscopy, complemented by band structure calculations. Our findings reveal that photoexcitation generates delocalized charge transfer (CT) states between Au orbitals in $[Au^{3+}Br_4]^-$ and $[Au^+Br_2]^-$ octahedra. These states are strongly coupled to specific phonon modes of the crystal, particularly the Au-Br stretching modes. The early-stage transient absorption dynamics are characterized by modulation of the absorption spectrum by coherent phonons, and a derivative-shaped feature attributed to band-edge renormalization. From this initial configuration, we reveal that a localized polaronic state formation occurs on a timescale of 10-20 ps, with a lattice relaxation energy of approximately 80 meV. This process is driven by strong electron-phonon coupling and a large displacement between potential energy surfaces of the localized state and the ground state.

These findings highlight the potential of $Cs_2Au_2Br_6$ as a lead-free perovskite material with unique electronic and vibrational properties. The possibility to control charge carrier localization and coherent lattice dynamics through electron-phonon coupling opens new avenues for designing materials with tailored optoelectronic properties. Engineering the crystal structure, for instance, by doping with another species may enable further tuning its electronic and vibrational properties.[57-59]

Even though the factors leading to slow vibrational dephasing in this system remain unclear, long phonon coherence times can enable control of non-equilibrium electronic configurations in solids.[60, 61] Employing ultrafast structural probes[62-64] like femtosecond X-ray scattering and electron diffraction could provide deeper insights into the mechanisms of vibrational relaxation and the role of coherent acoustic modes. A deeper understanding of charge and lattice dynamics is therefore critical for improving the efficiency and stability of next-generation optoelectronic devices.

## Methods

Materials:

Cesium bromide (CsBr, 99.999%, trace metals basis), Gold(III) bromide ($AuBr_3$, 99.9%, trace metals basis), γ-Butyrolactone (GBL, ≥ 99%, Reagent Plus) and Chlorobenzene (CB, anhydrous, 99.8%) were purchased from Sigma-Aldrich and were used without further purification.

Synthesis of $Cs_2Au_2Br_6$ Samples:

The $Cs_2Au_2Br_6$ thin film was fabricated on ITO substrate. The ITO substrates were ultrasonic cleaned with detergent, deionized water, ethanol and $O_2$ plasma asher for 15 min respectively. Then transfer the substances into a $N_2$-filled glove box. The precursor solution (0.4M CsBr and 0.4M $AuBr_3$ in GBL) was spin-coated at 5000 rpm (with a speed ramp of 2500 rpm s−1) for 45s, and 100 μL of chlorobenzene as the antisolvent was deposited onto the film at 20s. Subsequently, the samples were annealed at 150 °C for 15 min on a hotplate.

Steady-state Optical Measurements:

Absorption spectrum was obtained using an Agilent Technologies 8453 spectrophotometer. Emission spectrum was obtained on a Horiba Fluorolog-3 fluorimeter using a CCD detector, with an accumulation time of 10 seconds.

Density Functional Theory:

Density Functional Theory (DFT) calculations were conducted using the Projected Augmented Wave (PAW) method within the Vienna Ab initio Simulation Package (VASP).[65, 66] Vibrational modes and Raman intensities were computed employing the generalized gradient approximation with the Perdew, Burke, and Ernzerhof (PBE) parametrization to account for the exchange and correlation terms within the Kohn-Sham Hamiltonian,[67] including spin-orbit coupling (SOC). Plane waves were expanded up to an energy cutoff of 520 eV, with Brillouin zone sampling performed on a reciprocal grid of 4x4x2. It's worth highlighting that, in this context, precise convergence criteria and computationally demanding calculations are imperative due to the dependency of forces and Raman responses on the Hessian, representing the second derivative of nuclear positions. This requirement stands in contrast to electronic structure calculations, which typically converge prior to achieving detailed nuclear positions. To access the phonon modes and frequencies at the Γ point, we employed Density Functional Perturbation Theory (DFPT). For this phase, the structure underwent reoptimization with a force convergence target of 0.001 eV/Å and an energy convergence goal of $10^{-8}$ eV. We estimated the variations in the macroscopic dielectric tensor and obtained the relative Raman intensities by applying finite differences along each vibrational mode.[68] For electronic structure analysis, we turned to the hybrid functional of Heyd–Scuseria–Ernzerhof (HSE06)[69] to alleviate the effects of electron self-interaction and to prevent underestimation of band gaps. This involved another structural reoptimization, with a force convergence criterion set at 0.01 eV/Å and an energy convergence criterion of $10^{-6}$ eV. Phonon band structure and density of states are computed with Phononpy.[70]

Raman Spectroscopy:

Raman measurements were conducted by using a Renishaw InVia spectrometer with a frequency doubled Nd:YAG laser operating at 532 nm, a 2400 lines/mm grating, and Raman filter cutting 85 $cm^{-1}$ into the anti-Stokes part of the Raman spectra. The spectrometer focal length (250 mm), grating, and slit chosen gave a resolution of < 1 $cm^{-1}$ per pixel. The measurements were performed using a 50x objective and focusing the laser onto the sample. Complementary low intensity laser measurements were routinely performed to ensure that no laser induced heating would affect the spectra.

Transient Absorption Spectroscopy:

TA measurements were conducted by using a homemade femtosecond pump-probe setup. Laser pulses (8 W, 796 nm, 60 fs, 4 kHz) come out of Solstice (Spectra Physics) amplifier seeded by a femtosecond oscillator (Mai Tai SP, Spectra Physics). The laser output is split into two beams that each pump collinear optical parametric amplifiers (TOPAS-C, Light Conversion). The first one generates 500 nm wavelength pump pulses, while the other generates 1350 nm pulses that is focused onto a $CaF_2$ plate to generate a supercontinuum pulse which is used as the probe. The probe beam is delayed with respect to the pump using a mechanical delay stage. After supercontinuum generation, the probe beam is further divided into two parts: the former is focused on the sample overlapping with the pump pulse and the latter serving as a reference. After passing the sample, the probe beam is collimated again and relayed onto the entrance aperture of a prism spectrograph. The reference beam is directly relayed on the spectrograph. Both the measurement and reference beams are then dispersed onto a double photodiode array, each holding 512 elements (Pascher Instruments). The pump and probe beams are cross polarized with a Berek compensator in the path of the pump beam. A Glan–Thompson polarizer is placed after the sample to filter out excitation light scatter. Data analysis was performed with Origin 2024 and custom-written scripts in python.

## Acknowledgements:

We acknowledge financial support from Swedish Energy Agency (P2020-90219, P2020-90215) Swedish Research Council (2021-05207, 2023-05244) and KAW WACQT program. Collaboration with NanoLund is acknowledged. The computations were enabled by resources provided by the National Academic Infrastructure for Super-computing in Sweden (NAISS) via the projects NAISS 2023/5-276 and 2024/5-372. We are thankful to Dr Jens Uhlig for helpful discussions regarding data processing of transient absorption signals.

## References:

(1) Cheng, Y.; Peng, Y.; Jen, A. K.-Y.; Yip, H.-L. Development and challenges of metal halide perovskite solar modules. Solar RRL 2022, 6 (3), 2100545.

(2) Liu, X.-K.; Xu, W.; Bai, S.; Jin, Y.; Wang, J.; Friend, R. H.; Gao, F. Metal halide perovskites for light-emitting diodes. Nature Materials 2021, 20 (1), 10-21.

(3) Fu, J.; Ramesh, S.; Melvin Lim, J. W.; Sum, T. C. Carriers, quasi-particles, and collective excitations in halide perovskites. Chemical Reviews 2023, 123 (13), 8154-8231.

(4) Fu, J.; Xu, Q.; Han, G.; Wu, B.; Huan, C. H. A.; Leek, M. L.; Sum, T. C. Hot carrier cooling mechanisms in halide perovskites. Nature communications 2017, 8 (1), 1-9.

(5) Smith, M. D.; Jaffe, A.; Dohner, E. R.; Lindenberg, A. M.; Karunadasa, H. I. Structural origins of broadband emission from layered Pb–Br hybrid perovskites. Chemical science 2017, 8 (6), 4497-4504.

(6) Thouin, F.; Valverde-Chávez, D. A.; Quarti, C.; Cortecchia, D.; Bargigia, I.; Beljonne, D.; Petrozza, A.; Silva, C.; Srimath Kandada, A. R. Phonon coherences reveal the polaronic character of excitons in two-dimensional lead halide perovskites. Nature materials 2019, 18 (4), 349-356.

(7) Zheng, K.; Abdellah, M.; Zhu, Q.; Kong, Q.; Jennings, G.; Kurtz, C. A.; Messing, M. E.; Niu, Y.; Gosztola, D. J.; Al-Marri, M. J.; et al. Direct experimental evidence for photoinduced strong-coupling polarons in organolead halide perovskite nanoparticles. The Journal of Physical Chemistry Letters 2016, 7 (22), 4535-4539.

(8) He, Y.; Liu, S.; Yao, Z.; Zhao, Q.; Chabera, P.; Zheng, K.; Yang, B.; Pullerits, T. n.; Chen, J. Nature of self-trapped exciton emission in zero-dimensional $Cs_2ZrCl_6$ perovskite nanocrystals. The Journal of Physical Chemistry Letters 2023, 14 (34), 7665-7671.

(9) Suzuki, T.; Shinohara, Y.; Lu, Y.; Watanabe, M.; Xu, J.; Ishikawa, K. L.; Takagi, H.; Nohara, M.; Katayama, N.; Sawa, H.; et al. Detecting electron-phonon coupling during photoinduced phase transition. Physical Review B 2021, 103 (12), L121105.

(10) Lin, W.; Canton, S. E.; Zheng, K.; Pullerits, T. Carrier Cooling in Lead Halide Perovskites: A Perspective on Hot Carrier Solar Cells. ACS Energy Letters 2023, 9 (1), 298-307.

(11) Koshihara, S.; Ishikawa, T.; Okimoto, Y.; Onda, K.; Fukaya, R.; Hada, M.; Hayashi, Y.; Ishihara, S.; Luty, T. Challenges for developing photo-induced phase transition (PIPT) systems: from classical (incoherent) to quantum (coherent) control of PIPT dynamics. Physics Reports 2022, 942, 1-61.

(12) Guo, Q.; Zhao, X.; Song, B.; Luo, J.; Tang, J. Light emission of self-trapped excitons in inorganic metal halides for optoelectronic applications. Advanced Materials 2022, 34 (52), 2201008.

(13) Giovanni, D.; Righetto, M.; Zhang, Q.; Lim, J. W. M.; Ramesh, S.; Sum, T. C. Origins of the long-range exciton diffusion in perovskite nanocrystal films: photon recycling vs exciton hopping. Light: Science & Applications 2021, 10 (1), 2.

(14) Ramesh, S.; Giovanni, D.; Righetto, M.; Ye, S.; Fresch, E.; Wang, Y.; Collini, E.; Mathews, N.; Sum, T. C. Tailoring the energy manifold of quasi-two-dimensional perovskites for efficient carrier extraction. Advanced Energy Materials 2022, 12 (10), 2103556.

(15) Ponseca Jr, C. S.; Savenije, T. J.; Abdellah, M.; Zheng, K.; Yartsev, A.; Pascher, T. r.; Harlang, T.; Chabera, P.; Pullerits, T.; Stepanov, A. Organometal halide perovskite solar cell materials rationalized: ultrafast charge generation, high and microsecond-long balanced mobilities, and slow recombination. Journal of the American Chemical Society 2014, 136 (14), 5189-5192.

(16) Ji, F.; Boschloo, G.; Wang, F.; Gao, F. Challenges and progress in lead-free halide double perovskite solar cells. Solar RRL 2023, 7 (6), 2201112.

(17) Yan, Y.; Pullerits, T.; Zheng, K.; Liang, Z. Advancing tin halide perovskites: strategies toward the ASnX3 paradigm for efficient and durable optoelectronics. ACS Energy Letters 2020, 5 (6), 2052-2086.

(18) McCall, K. M.; Stoumpos, C. C.; Kostina, S. S.; Kanatzidis, M. G.; Wessels, B. W. Strong electron–phonon coupling and self-trapped excitons in the defect halide perovskites A3M2I9 (A= Cs, Rb; M= Bi, Sb). Chemistry of Materials 2017, 29 (9), 4129-4145.

(19) Wright, A. D.; Buizza, L. R.; Savill, K. J.; Longo, G.; Snaith, H. J.; Johnston, M. B.; Herz, L. M. Ultrafast excited-state localization in Cs2AgBiBr6 double perovskite. The journal of physical chemistry letters 2021, 12 (13), 3352-3360.

(20) Wu, B.; Ning, W.; Xu, Q.; Manjappa, M.; Feng, M.; Ye, S.; Fu, J.; Lie, S.; Yin, T.; Wang, F.; et al. Strong self-trapping by deformation potential limits photovoltaic performance in bismuth double perovskite. Science Advances 2021, 7 (8), eabd3160.

(21) Ji, F.; Klarbring, J.; Zhang, B.; Wang, F.; Wang, L.; Miao, X.; Ning, W.; Zhang, M.; Cai, X.; Bakhit, B.; et al. Remarkable Thermochromism in the Double Perovskite Cs2NaFeCl6. Advanced Optical Materials 2024, 12 (8), 2301102.

(22) Luo, J.; Wang, X.; Li, S.; Liu, J.; Guo, Y.; Niu, G.; Yao, L.; Fu, Y.; Gao, L.; Dong, Q.; et al. Efficient and stable emission of warm-white light from lead-free halide double perovskites. Nature 2018, 563 (7732), 541-545.

(23) Pan, W.; Wu, H.; Luo, J.; Deng, Z.; Ge, C.; Chen, C.; Jiang, X.; Yin, W.-J.; Niu, G.; Zhu, L.; et al. Cs2AgBiBr6 single-crystal X-ray detectors with a low detection limit. Nature photonics 2017, 11 (11), 726-732.

(24) Heo, J.; Yu, L.; Altschul, E.; Waters, B. E.; Wager, J. F.; Zunger, A.; Keszler, D. A. Cutas3: intermetal d–d transitions enable high solar absorption. Chemistry of Materials 2017, 29 (6), 2594-2598.

(25) Xiao, Z.; Meng, W.; Wang, J.; Mitzi, D. B.; Yan, Y. Searching for promising new perovskite-based photovoltaic absorbers: the importance of electronic dimensionality. Materials Horizons 2017, 4 (2), 206-216.

(26) Yang, B.; Hong, F.; Chen, J.; Tang, Y.; Yang, L.; Sang, Y.; Xia, X.; Guo, J.; He, H.; Yang, S.; et al. Colloidal synthesis and charge-carrier dynamics of Cs2AgSb1− yBiyX6 (X: Br, Cl; 0≤ y≤ 1) double perovskite nanocrystals. Angewandte Chemie 2019, 131 (8), 2300-2305.

(27) Buizza, L. R.; Wright, A. D.; Longo, G.; Sansom, H. C.; Xia, C. Q.; Rosseinsky, M. J.; Johnston, M. B.; Snaith, H. J.; Herz, L. M. Charge-carrier mobility and localization in semiconducting Cu2AgBiI6 for photovoltaic applications. ACS energy letters 2021, 6 (5), 1729-1739.

(28) Debbichi, L.; Lee, S.; Cho, H.; Rappe, A. M.; Hong, K. H.; Jang, M. S.; Kim, H. Mixed valence perovskite Cs2Au2I6: a potential material for thin-film Pb-free photovoltaic cells with ultrahigh efficiency. Advanced Materials 2018, 30 (12), 1707001.

(29) Kojima, N.; Kitagawa, H. Optical investigation of the intervalence charge-transfer interactions in the three-dimensional gold mixed-valence compounds Cs2Au2X6 (X= Cl, Br or I). Journal of the Chemical Society, Dalton Transactions 1994, (3), 327-331.

(30) Liu, X.; Moritomo, Y.; Ichida, M.; Nakamura, A.; Kojima, N. Photoinduced phase transition in a mixed-valence gold complex. Physical Review B 2000, 61 (1), 20.

(31) Son, J.-Y.; Mizokawa, T.; Quilty, J.; Takubo, K.; Ikeda, K.; Kojima, N. Photoinduced valence transition in gold complexes Cs2Au2X6 (X= Cl and Br) probed by x-ray photoemission spectroscopy. Physical Review B 2005, 72 (23), 235105.

(32) Naumov, P.; Huangfu, S.; Wu, X.; Schilling, A.; Thomale, R.; Felser, C.; Medvedev, S.; Jeschke, H. O.; Von Rohr, F. O. Large resistivity reduction in mixed-valent CsAuBr3 under pressure. Physical Review B 2019, 100 (15), 155113.

(33) Lindquist, K. P.; Eghdami, A.; Deschene, C. R.; Heyer, A. J.; Wen, J.; Smith, A. G.; Solomon, E. I.; Lee, Y. S.; Neaton, J. B.; Ryan, D. H.; et al. Stabilizing Au2+ in a mixed-valence 3D halide perovskite. Nature Chemistry 2023, 15 (12), 1780-1786.

(34) Kangsabanik, J.; Ghorui, S.; Aslam, M.; Alam, A. Optoelectronic Properties and Defect Physics of Lead-Free Photovoltaic Absorbers Cs2AuIAuIIIX6 (X= I, Br). Physical Review Applied 2020, 13 (1), 014005.

(35) Suzuki, S.; Tsuyama, M. Theoretical study on electronic and optical properties of mixed valence perovskite Cs2Au2X6 (X= Cl, Br, I). Japanese Journal of Applied Physics 2019, 58 (11), 111002.

(36) Liu, X.; Matsuda, K.; Moritomo, Y.; Nakamura, A.; Kojima, N. Electronic structure of the gold complexes Cs2Au2X6 (X= I, Br, and Cl). Physical Review B 1999, 59 (12), 7925.

(37) Fu, J.; Li, M.; Solanki, A.; Xu, Q.; Lekina, Y.; Ramesh, S.; Shen, Z. X.; Sum, T. C. Electronic states modulation by coherent optical phonons in 2D halide perovskites. Advanced Materials 2021, 33 (11), 2006233.

(38) Garrett, G. A.; Albrecht, T.; Whitaker, J.; Merlin, R. Coherent THz phonons driven by light pulses and the Sb problem: what is the mechanism? Physical review letters 1996, 77 (17), 3661.

(39) Ramesh, S.; Feng, M.; Furuhashi, T.; Sum, T. C. Enhancing Two-Dimensional Electronic Spectroscopy for Layered Halide Perovskites. ACS Photonics 2023, 10 (12), 4456-4464.

(40) Dhar, L.; Rogers, J. A.; Nelson, K. A. Time-resolved vibrational spectroscopy in the impulsive limit. Chemical Reviews 1994, 94 (1), 157-193.

(41) Klemens, P. Anharmonic decay of optical phonons. Physical Review 1966, 148 (2), 845.

(42) Mante, P.-A.; Stoumpos, C. C.; Kanatzidis, M. G.; Yartsev, A. Electron–acoustic phonon coupling in single crystal $CH_3NH_3PbI_3$ perovskites revealed by coherent acoustic phonons. Nature communications 2017, 8 (1), 14398.

(43) Stevens, T.; Kuhl, J.; Merlin, R. Coherent phonon generation and the two stimulated Raman tensors. Physical Review B 2002, 65 (14), 144304.

(44) Bragas, A. V.; Aku-Leh, C.; Costantino, S.; Ingale, A.; Zhao, J.; Merlin, R. Ultrafast optical generation of coherent phonons in $CdTe_{1-x}Se_x$ quantum dots. Physical Review B 2004, 69 (20), 205306.

(45) Coropceanu, V.; Cornil, J.; da Silva Filho, D. A.; Olivier, Y.; Silbey, R.; Brédas, J.-L. Charge transport in organic semiconductors. Chemical reviews 2007, 107 (4), 926-952.

(46) Pollard, W.; Fragnito, H.; Bigot, J.-Y.; Shank, C.; Mathies, R. A. Quantum-mechanical theory for 6 fs dynamic absorption spectroscopy and its application to nile blue. Chemical physics letters 1990, 168 (3-4), 239-245.

(47) Sagar, D.; Cooney, R. R.; Sewall, S. L.; Dias, E. A.; Barsan, M. M.; Butler, I. S.; Kambhampati, P. Size dependent, state-resolved studies of exciton-phonon couplings in strongly confined semiconductor quantum dots. Physical Review B 2008, 77 (23), 235321.

(48) Ghosh, T.; Aharon, S.; Etgar, L.; Ruhman, S. Free carrier emergence and onset of electron–phonon coupling in methylammonium lead halide perovskite films. Journal of the American Chemical Society 2017, 139 (50), 18262-18270.

(49) Biswas, S.; Zhao, R.; Alowa, F.; Zacharias, M.; Sharifzadeh, S.; Coker, D. F.; Seferos, D. S.; Scholes, G. D. Exciton polaron formation and hot-carrier relaxation in rigid Dion–Jacobson-type two-dimensional perovskites. Nature Materials 2024, 1-7.

(50) Dasgupta, J.; Tomar, D.; Deshpande, S.; Gupta, S.; Nayak, P.; Chakrabarty, S. Exciton-Polaron in a quasi-one-dimensional chain of hexyl-diammonium-BiI5 octahedra. 2024.

(51) Haug, H.; Schmitt-Rink, S. Basic mechanisms of the optical nonlinearities of semiconductors near the band edge. JOSA B 1985, 2 (7), 1135-1142.

(52) Wu, X.; Trinh, M. T.; Zhu, X.-Y. Excitonic many-body interactions in two-dimensional lead iodide perovskite quantum wells. The Journal of Physical Chemistry C 2015, 119 (26), 14714-14721.

(53) Parson, W. W. Modern optical spectroscopy; Springer, 2007.

(54) Schulze, J.; Torbjörnsson, M.; Kühn, O.; Pullerits, T. Exciton coupling induces vibronic hyperchromism in light-harvesting complexes. New Journal of Physics 2014, 16 (4), 045010.

(55) Zelewski, S.; Urban, J. M.; Surrente, A.; Maude, D. K.; Kuc, A.; Schade, L.; Johnson, R.; Dollmann, M.; Nayak, P.; Snaith, H.; et al. Revealing the nature of photoluminescence emission in the metal-halide double perovskite $Cs_2AgBiBr_6$. Journal of Materials Chemistry C 2019, 7 (27), 8350-8356.

(56) Zeng, R.; Zhang, L.; Xue, Y.; Ke, B.; Zhao, Z.; Huang, D.; Wei, Q.; Zhou, W.; Zou, B. Highly efficient blue emission from self-trapped excitons in stable $Sb^{3+}$-doped $Cs_2NaInCl_6$ double perovskites. The Journal of Physical Chemistry Letters 2020, 11 (6), 2053-2061.

(57) Rachna; Singh, A.; Kumar, S.; Sapra, S. Ultrafast Dynamics of Self-Trapped Excitons in $Cs_2AgInCl_6$: $Al^{3+}$ Double Perovskite Nanocrystals. Nano Letters 2024.

(58) Manna, D.; Kangsabanik, J.; Das, T. K.; Das, D.; Alam, A.; Yella, A. Lattice Dynamics and Electron–Phonon Coupling in Lead-Free $Cs_2AgIn_{1-x}Bi_xCl_6$ Double Perovskite Nanocrystals. The Journal of Physical Chemistry Letters 2020, 11 (6), 2113-2120.

(59) Siddique, H.; Xu, Z.; Li, X.; Saeed, S.; Liang, W.; Wang, X.; Gao, C.; Dai, R.; Wang, Z.; Zhang, Z. Anomalous Octahedron Distortion of Bi-Alloyed $Cs_2AgInCl_6$ Crystal via XRD, Raman, Huang–Rhys Factor, and Photoluminescence. The Journal of Physical Chemistry Letters 2020, 11 (22), 9572-9578.

(60) Teitelbaum, S. W.; Shin, T.; Wolfson, J. W.; Cheng, Y.-H.; Molesky, I. J.; Kandyla, M.; Nelson, K. A. Real-time observation of a coherent lattice transformation into a high-symmetry phase. Physical Review X 2018, 8 (3), 031081.

(61) Mante, P.-A.; Ong, C. S.; Shapiro, D. F.; Yartsev, A.; Grånäs, O.; Eriksson, O. Photo-induced Hidden Phase of 1T-TaS2 with Tunable Lifetime. arXiv preprint arXiv:2203.13509 2022.

(62) Chergui, M.; Collet, E. Photoinduced structural dynamics of molecular systems mapped by time-resolved X-ray methods. Chemical reviews 2017, 117 (16), 11025-11065.

(63) Seiler, H.; Zahn, D.; Taylor, V. C.; Bodnarchuk, M. I.; Windsor, Y. W.; Kovalenko, M. V.; Ernstorfer, R. Direct observation of ultrafast lattice distortions during exciton–polaron formation in lead halide perovskite nanocrystals. ACS nano 2023, 17 (3), 1979-1988.

(64) Liu, C.; Wang, Y.; Geng, H.; Zhu, T.; Ertekin, E.; Gosztola, D.; Yang, S.; Huang, J.; Yang, B.; Han, K.; et al. Asynchronous photoexcited electronic and structural relaxation in lead-free perovskites. Journal of the American Chemical Society 2019, 141 (33), 13074-13080.

(65) Kresse, G.; Furthmüller, J. Efficient iterative schemes for ab initio total-energy calculations using a plane-wave basis set. Physical review B 1996, 54 (16), 11169.

(66) Kresse, G.; Joubert, D. From ultrasoft pseudopotentials to the projector augmented-wave method. Physical review B 1999, 59 (3), 1758.

(67) Perdew, J. P.; Burke, K.; Ernzerhof, M. Generalized gradient approximation made simple. Physical review letters 1996, 77 (18), 3865.

(68) Fonari, A.; Stauffer, S. vasp_raman.py; https://github.com/raman-sc/VASP/, 2013.

(69) Krukau, A. V.; Vydrov, O. A.; Izmaylov, A. F.; Scuseria, G. E. Influence of the exchange screening parameter on the performance of screened hybrid functionals. The Journal of chemical physics 2006, 125 (22).

(70) Togo, A.; Chaput, L.; Tadano, T.; Tanaka, I. Implementation strategies in phonopy and phono3py. Journal of Physics: Condensed Matter 2023, 35 (35), 353001.

# Supplementary Information

Coherent Phonons, Localization and Slow Polaron Formation in Lead-free Gold Perovskite

Sankaran Ramesh[1], Yonghong Wang[2], Pavel Chabera[1], Rafael Araujo[3], Mustafa Aboulsaad[3], Tomas Edvinsson[3], Feng Gao[2], Tönu Pullerits[1]

[1] Division of Chemical Physics and NanoLund, Lund University, Box 124, 221 00 Lund, Sweden

[2] Department of Physics, Chemistry, and Biology (IFM), Linköping University, Linköping, 581 83, Sweden

[3] Department of Materials Science and Engineering – Solid State Physics, Uppsala University, Box 534, SE-75121 Uppsala, Sweden

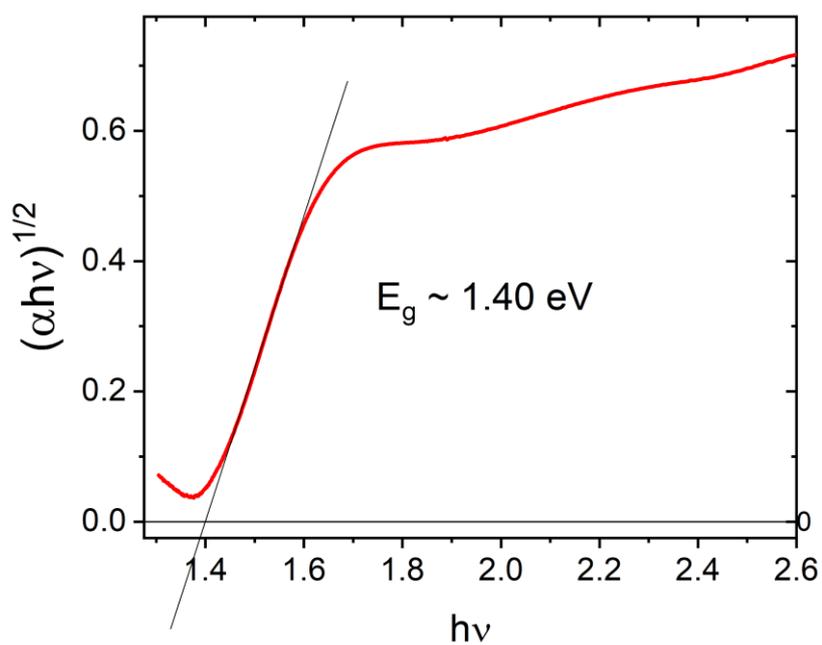

Figure S 1: Tauc Plot from Steady-State absorption spectrum of Cs$_2$Au$_2$Br$_6$

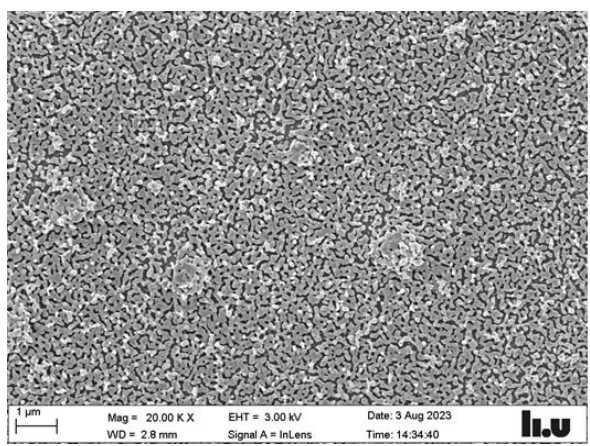 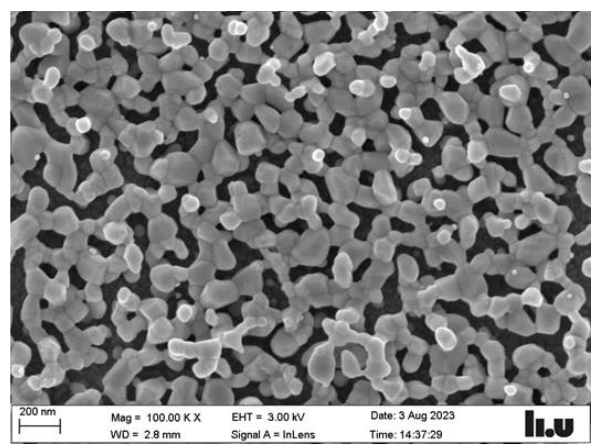

Figure S 2: SEM images of Cs$_2$Au$_2$Br$_6$ at different magnifications

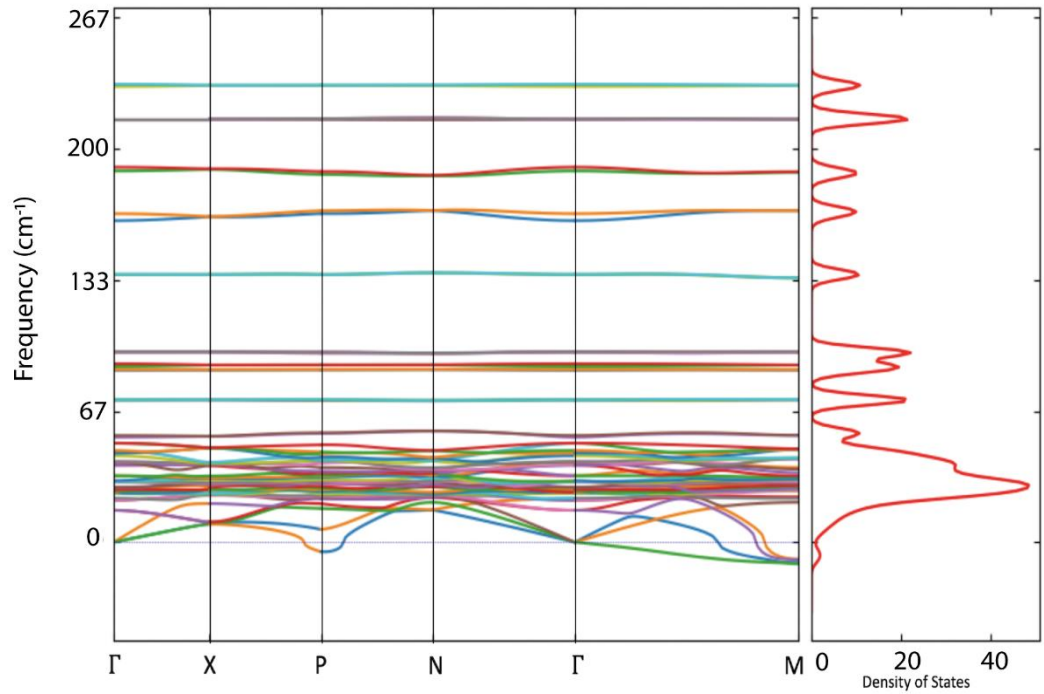

Figure S 3: Calculated Phonon bandstructure of $Cs_2Au_2Br_6$ with density of states.

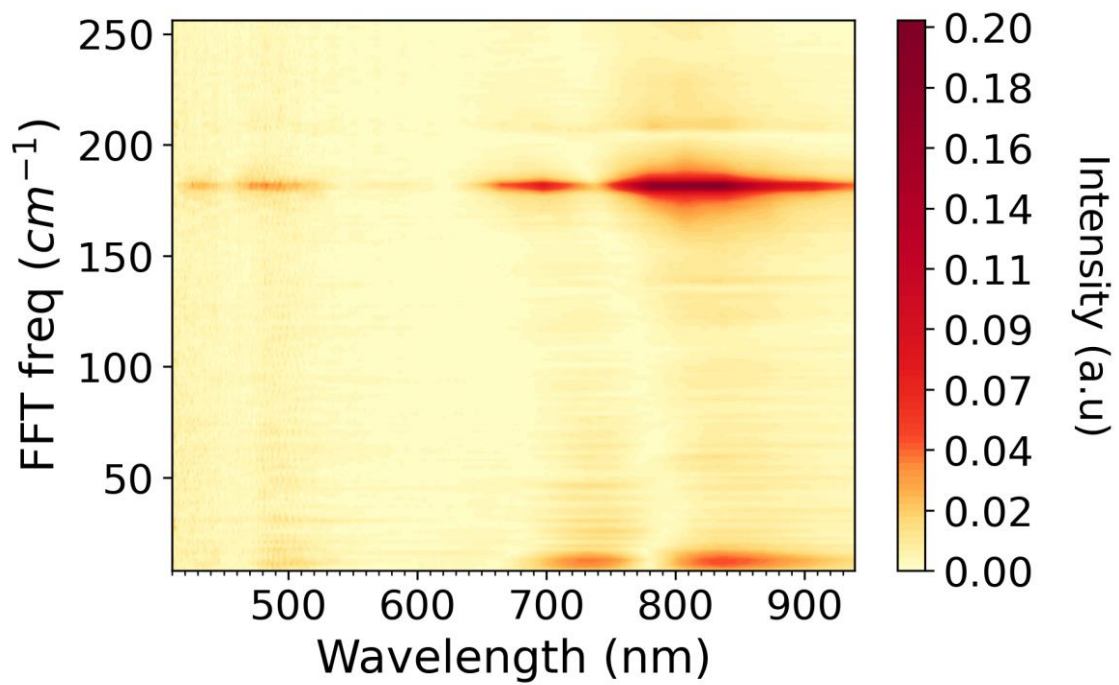

Figure S 4: Fourier transformed residual fs-TA spectrum of Cs2Au2Br6.

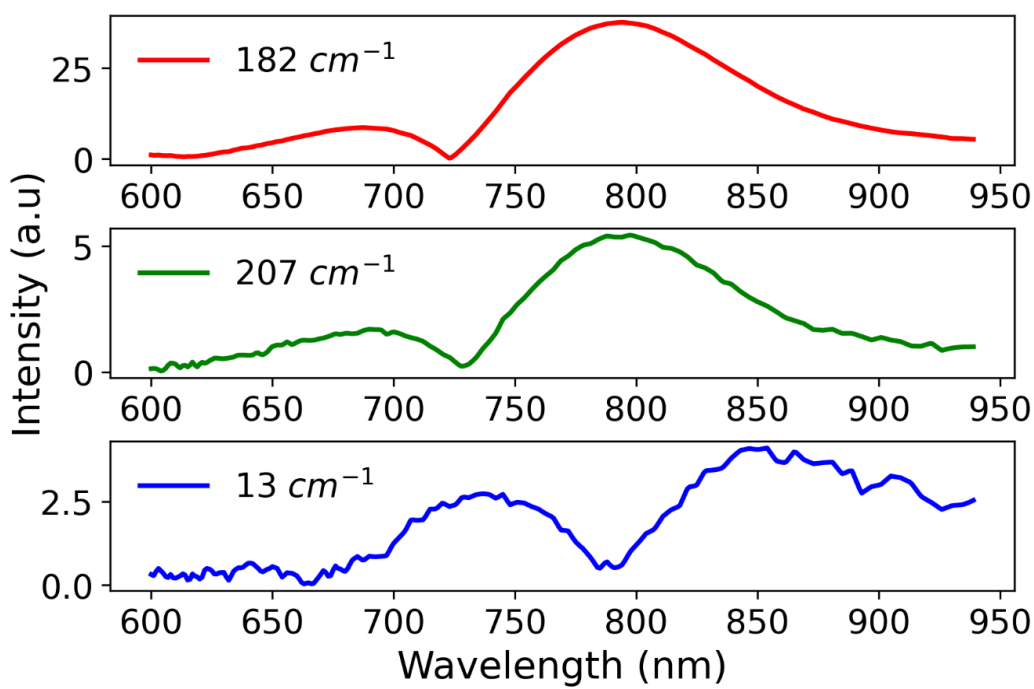

*Figure S 5: Slices of Fourier transformed residual TA spectrum at M1, M2 and M3 phonon frequencies.*

## Supplementary Note 1: Optimized Crystal Structure of $Cs_2Au_2Br_6$:

In the crystal structure under consideration, gold atoms are distributed across two distinct sites and are coordinated by six bromine atoms, forming octahedral structures. These octahedra share bromine atoms with one another and are arranged alternately, creating a structure that alternates between compressed and elongated configurations, resulting in [AuBr2] and [AuBr4] units. Within this framework, the gold atoms exhibit varying oxidation states. Upon performing structural energy minimization with a hybrid functional, the structure exhibited a volumetric expansion of approximately 7% in comparison to the experimental value. This expansion is primarily attributed to variations in the lattice parameters along the a and b axes. For the compressed octahedra, the gold-bromine bond distances are measured at 2.41 Å and 3.22 Å, while the corresponding experimental values are 2.39 Å and 3.05 Å. In the case of the elongated octahedra, the bond lengths are 2.45 Å and 3.27 Å, whereas the experimental values are 2.44 Å and 3.25 Å. Notably, the longer bonds in the ab-plane of the compressed octahedra ([AuBr4]) account for the observed volumetric expansion in the calculations when compared to the experimental data.

## Supplementary Note 2: Electronic Band Structure of $Cs_2Au_2Br_6$:

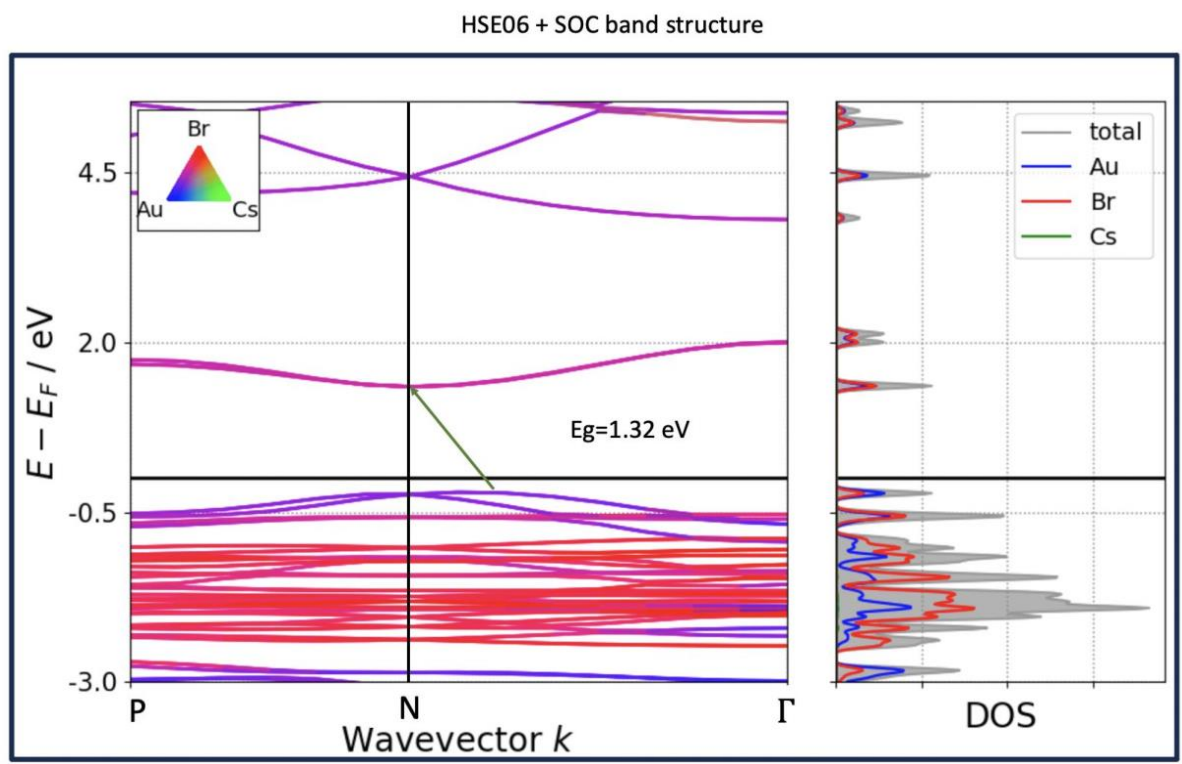

Figure S 6: Projected band structure in the atoms and Projected density of states (PDOS). In the band structure plot, a color map is used to represent the level of mixing or hybridization of the reference band. For the PDOS figure, gray background represents the total density of states while the blue is the gold projected DOS, red is the Br projected DOS and green is the Cs projected DOS.

The band structure analysis of $Cs_2Au_2Br_6$ has revealed an indirect band gap of 1.32 eV. The top of the valence band is located at the K point [0.0 0.42 0.00] (from N to $\Gamma$ direction), while the bottom of the conduction band is at the N point. It's worth noting that the difference between the direct and indirect band gaps is minor, as transitioning from N to N yields a band gap of 1.35 eV. The partial density of states (PDOS) and band structure indicate that the top of the valence band is predominantly formed by gold states, albeit with some contribution from bromine states (Figure S 6). Conversely, the bottom of the valence band mainly comprises bromine states but still exhibits some contribution from gold states. The PDOS shows contributions from states of both bromine and gold in the energy range from 0 eV to -3 eV, signifying the hybridization of p and d orbitals between gold and bromine. Examining the PDOS of the d states of gold atoms (Figure S 7) reveals that the gold coordinated in the compressed form (Au2) octahedra exhibits more contribution close to the Fermi level than its counterpart (Au1 - elongated), which has states shifted to lower energies. This indicates a higher level of hybridization between gold d states and bromine p states in the elongated octahedra, which is expected due to the coordination of four bromine atoms in contrast to two in the compressed octahedra.

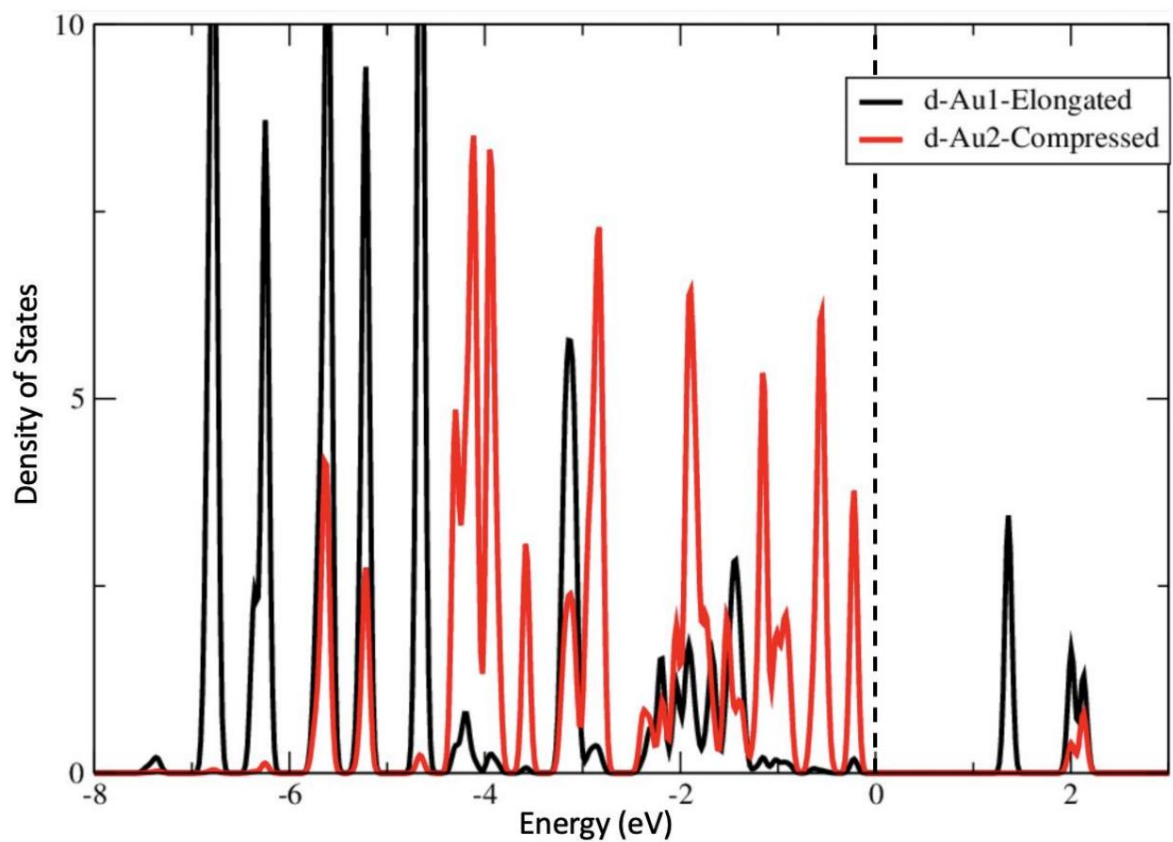

Figure S 7: rojected density of states (PDOS) on the d states of two Au atoms (Au1 elongated and Au2 compressed).

Bader charge analysis was conducted to determine the oxidation states of each element in $Cs_2Au_2Br_6$. As expected, caesium (Cs) exhibits an average charge of 0.89. In the case of bromides, two different Bader charges are observed, with values of -0.46 and -0.51. The bromides with the higher Bader charge are those sharing corners between the octahedra in the c direction. For gold, the Bader charges are +0.75 and +0.34, with Au1 (elongated) having the higher Bader charge and Au2 (compressed) having the lower Bader charge. While the total ionic bond picture would suggest Br to have a -1 atomic charge, the effects of bond hybridization, as evidenced in the Density of States (DOS), and the inherent self-interaction error in DFT calculations lead to effective charges of -0.46 and -0.51 for bromine. Similarly, the total ionic picture of gold as +1 and +3 is not observed. Nevertheless, the Bader charges for Au1 and Au2 differ and align with the expected trend of Au1 being more oxidized than Au2.

## Supplementary Note 3: Vibrational Properties and Raman-active Modes:

The phonon band structure plot for $Cs_2Au_2Br_6$ has been generated, and it indicates the absence of modes with negative frequencies or only very minor such modes. This observation confirms the dynamical stability of the structure under consideration.

Raman intensities have been computed for $Cs_2Au_2Br_6$, revealing four Raman active modes at specific frequencies: 191 cm$^{-1}$, 163 cm$^{-1}$, 136 cm$^{-1}$, and 89 cm$^{-1}$. The Raman active mode with the highest frequency, 191 cm$^{-1}$, is associated with the stretching of Au-Br bonds in the compressed octahedra. In contrast, the mode with the highest intensity is linked to the stretching of Br-Au bonds in the elongated octahedra. The other two active modes are also related to the vibrational behaviour of the elongated octahedra, as illustrated in Figure 5. These Raman active modes and their associated vibrations provide valuable insights into the dynamic behavior and structural properties of $Cs_2Au_2Br_6$, enhancing our understanding of its physical characteristics.

## Supplementary Note 4: Impulsive Stimulated Raman Scattering and Lattice Reorganization Energy:

Impulsive Stimulated Raman scattering (ISRS) is a mechanism by which the material interacts with a short-duration, broadband optical pulse.[1] It involves coherent interaction of the system with two components of the pulse separated in frequency by the phonon energy $\Omega$, generating coherent lattice vibrations that in turn, modulate the electronic energy. Considering stimulated Raman scattering of the incident light pulse by the coupled electron-lattice system, the phonon amplitude Q can be described using the driven oscillator equation:

$$\frac{d^2Q}{dt^2} + 2\Gamma \frac{dQ}{dt} + \Omega Q = F/\mu \qquad (S\ 1)$$

Where $\Gamma$ and $\mu$ are the damping constant and reduced mass, respectively. Considering a short-duration laser pulse incident on the sample along the z - direction and that the electric field of the pulse is E, The driving force for the oscillations F(z) is given by:[2-4]

$$F(z, \Omega) = \frac{1}{8\pi^2} \int \int_{-\infty}^{\infty} E(z, \omega)\, E^*(z, \omega - \Omega) R_Q(\omega, \omega - \Omega)\, d\omega \qquad (S\ 2)$$

$R_Q(\omega, \omega - \Omega)$ is the element of the Raman susceptibility tensor which determines the Raman scattering strength for the phonon mode with frequency $\Omega$. For the excitation pulse with a duration shorter than $(1/\Gamma)$, the Raman susceptibility can be written for $\Gamma \to 0$ in terms of the complex dielectric constant $(\varepsilon_{Re} + i\varepsilon_{Im})$ as:

$$R_Q(\omega, \omega - \Omega) = C \, \Xi \left[ \frac{d\varepsilon_{Re}}{d\omega} + 2i \frac{\varepsilon_{Im}}{\Omega} \right] \tag{S 3}$$

Where $\Xi$ is the deformation potential which is the change in electronic energy $E_m$ with change in nuclear displacement and C is a constant. For an electron at a lattice site $m$ interacting with a phonon mode with a nuclear co-ordinate $q_m$, it is defined as:

$$\Xi = \frac{\partial E_m}{\partial q_m} \tag{S 4}$$

Momentum conservation dictates that the scattered phonon wave vector $q \sim 0$, hence the deformation potential which measures the electron-phonon coupling can be considered a constant for coherent optical phonons modes that have negligible dispersion at $q = 0$. The phonon mode amplitude $Q(t)$ with a phenomenological dephasing rate $\Gamma$ is then given by: [5]

$$Q(t) = \frac{C\varepsilon_{Im}}{\Omega_1^2} \Xi \, I_0 \, e^{-\Gamma t} \cos(\Omega_1 t + \theta) \tag{S 5}$$

Where C is a constant pre-factor, $\Omega_1 = \sqrt{\Omega^2 - \Gamma^2}$, $\theta = \tan^{-1}\left( \frac{d\varepsilon_{Re}}{d\omega} \Big/ \frac{2\varepsilon_{Im}}{\Omega_1} \right)$ and $I_0 = \int_{-\infty}^{\infty} |E(t)|^2 dt$, the integrated pulse intensity. The phonon amplitude is linearly proportional to the incident pulse intensity and the strength of electron-phonon coupling. The lattice vibration affects the differential absorption via spontaneous Raman scattering of the probe electric field to generate the oscillatory signal: [4, 5]

$$\Delta A_{osc} = \frac{1}{\ln 10} \frac{\omega_0 L}{cn} \frac{d\varepsilon_{Im}}{dE} \Xi \, Q \tag{S 6}$$

Where L is the sample thickness, n is the refractive index, and c is the speed of light in vacuum. Using Equation ( 3 ) and (S 6) ,

$$\Delta A_{osc} \propto \frac{d\varepsilon_{\text{Im}}}{dE} \frac{\Xi^2}{\Omega_1^2} I_0 \qquad (S\ 7)$$

Considering a single charge in the lattice interacting with a phonon mode M1, the coherent phonons modulate its energy and subsequently the lattice undergoes relaxation resulting in a localized polaron state. This process can be understood using displaced potential energy surfaces, as discussed in Figure 5c-d in the main text. The lattice reorganization energy $\lambda$ due to localization on the potential energy surface of the mode with frequency $\Omega$ is related to the deformation potential constant $\Xi$ as follows: [6]

$$\lambda = \frac{\Xi^2}{2m\Omega^2} \qquad (S\ 8)$$

Where $m$ is the reduced mass of the phonon mode. Comparing this result with Equation (S 7),

$$\lambda \propto \frac{\Delta A_{osc}}{\left(\frac{d\varepsilon_{\text{Im}}}{dE}\right)} \qquad (S\ 9)$$

The derivative $\frac{d\varepsilon_{\text{Im}}}{dE}$ is proportional to the derivative of the absorption spectrum $\left(\frac{dA_0}{dE}\right)$. This can be used with Equation (S 9) to obtain the model in the main text (Equation (6) in the main text) to estimate the polaron binding energy and Huang-Rhys factors.

## Supplementary Note 5: Calculation of the Imaginary Dielectric Constant

The complex dielectric constant of a medium $\varepsilon$ is defined as $\varepsilon = \varepsilon_{\text{Re}} + i\varepsilon_{\text{Im}}$, where $\varepsilon_{\text{Re}}$ and $\varepsilon_{\text{Im}}$ are the real and imaginary parts. They are related to the complex refractive index as:

$$\varepsilon_{\text{Re}} + i\varepsilon_{\text{Im}} = (n + i\kappa)^2 \qquad (S\ 10)$$

Where $n$ and $\kappa$ are the real part of the refractive index and the extinction coefficient. The latter is related to the optical absorption coefficient α as:

$$\alpha = 4\pi\kappa/\lambda \qquad (S\ 11)$$

The imaginary part of dielectric constant $\varepsilon_{\text{Im}}$ can be written as $\varepsilon_{\text{Im}} = 2n\kappa$, and the real part of refractive index $n$ can be obtained from $\kappa$ using the Kramers-Kronig relations:

$$n(\omega) = 1 + \frac{2}{\pi}\, P \int_0^\infty \frac{\Omega\, \kappa(\Omega)}{\Omega^2 - \omega^2}\, d\Omega \qquad (S\ 12)$$

Where P is the Cauchy Principal Value. The plot at the bottom of Figure 4 (a) in the main text is obtained from the steady-state absorption spectrum of $Cs_2Au_2Br_6$ by using Equation (S 12).

## Supplementary Note 6: Derivate-shaped Feature at Band-edge:

The derivative-shaped TA spectrum ΔA at the band-edge is fit using a function of two gaussian peaks:

$$\Delta A = \sum_{i=1,2} C_i\, exp\left(\frac{(x-\mu_i)^2}{\sigma_i^2}\right) \qquad (S\ 13)$$

Where $C, \mu, \sigma$ are the amplitude, center and standard deviation of the gaussian. The index $i = 1,2$ corresponds to the GSB and ESA peaks, respectively. The TA spectrum was fitted for all delays. Figure S 8(a) shows slices of the TA spectrum and their fits (magenta lines). Figure S 8 (b-d) shows the dynamics of $C, \mu, \sigma$ as a function of the delay time. The peak shows a blue shift within the first few ps, which is due to the disappearance of band-edge renormalization and broadening. At the later stage, the peak undergoes a substantial red shift as the localized state dominates the response.

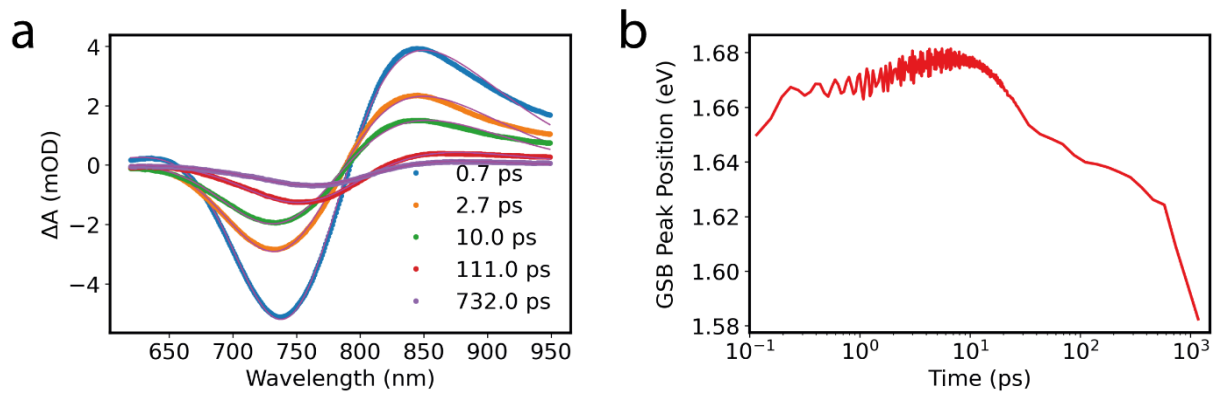

Figure S 8: Fitting of band-edge TA spectrum with two gaussian functions

## Supplementary Note 7: Global Analysis

Using singular value decomposition, Global analysis of the TA results at excitation wavelengths of 500 nm (above band-edge) and 670 nm (close to band-edge) is performed. The obtained decay-associated spectra and the associated lifetimes are shown in Figure S 9a-b.

Between the datasets in (a) and (b), the fitted components look similar in shape. The components are analysed as follows: The fastest component has a lifetime of 0.30 ps for excitation above the band-edge and 0.13 ps for excitation close to the band-edge. From the faster timescale of this component when excited close to the band-edge, we assign this to the thermalization of carriers to the band-edge CT state.

The second and third components have similar lifetimes between both datasets, of ~1.7 ps and ~23 ps. Since the two features are insensitive to the excess energy, we propose that they are related to the charge-lattice interactions. The 1.7 ps component is assigned to decay of the delocalized CT state. The 23 ps component consists of a negative and a positive feature, which is the signature of a shifting peak. Figure 6c in the main text shows that the ground state is the same for the initial excitation and the localized state. Hence, the ground state bleach signal will remain at the same energy after the polaronic state is formed. An alternate more probable scenario is that the relaxation can lead to shift of an excited state absorption (ESA) signal to higher energies. We conclude that a blue-shift of the ESA leads to disappearance of the initial positive band near the band-edge and the apparent red-shift of the negative band.

This timescale also corresponds to the dephasing time of coherent phonons in the experiment, which indicates vibrational relaxation that is as the polaron forms. Therefore, we assign this component to the lattice relaxation and polaron formation.

Finally, the slow component with an average lifetime of 1-2 ns, whose peak is red-shifted with respect to the band-edge CT state, is assigned to the recombination of the polaronic state.

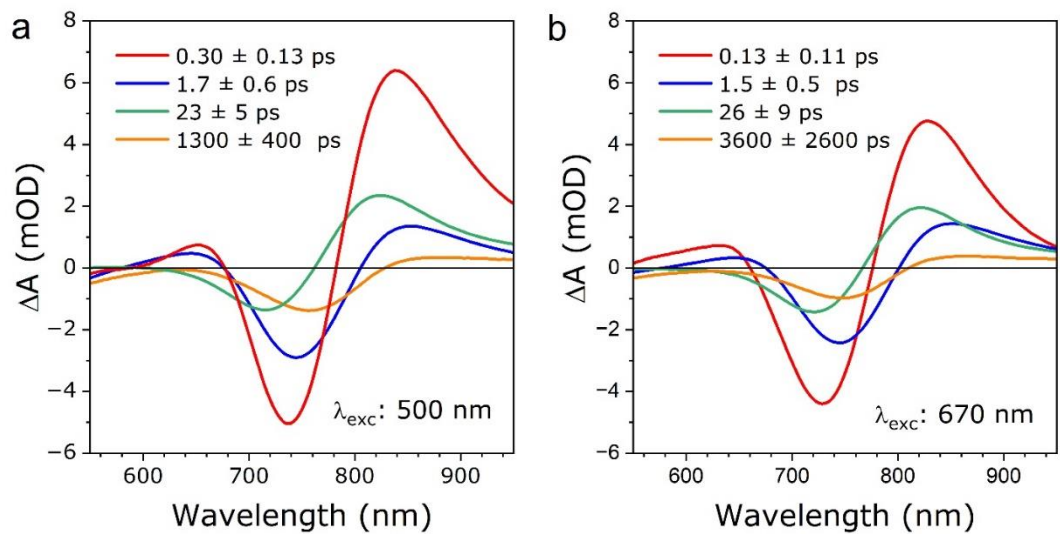

Figure S 9 : Global analysis of Band-edge TA signal for measurement at different excitation wavelengths (a) 500 nm and (b) 670 nm

# References

(1) Dhar, L.; Rogers, J. A.; Nelson, K. A. Time-resolved vibrational spectroscopy in the impulsive limit. Chemical Reviews 1994, 94 (1), 157-193.

(2) Garrett, G. A.; Albrecht, T.; Whitaker, J.; Merlin, R. Coherent THz phonons driven by light pulses and the Sb problem: what is the mechanism? Physical review letters 1996, 77 (17), 3661.

(3) Stevens, T.; Kuhl, J.; Merlin, R. Coherent phonon generation and the two stimulated Raman tensors. Physical Review B 2002, 65 (14), 144304.

(4) Bragas, A. V.; Aku-Leh, C.; Costantino, S.; Ingale, A.; Zhao, J.; Merlin, R. Ultrafast optical generation of coherent phonons in CdTe $_{1-x}$ Se $_x$ quantum dots. Physical Review B 2004, 69 (20), 205306.

(5) Fu, J.; Li, M.; Solanki, A.; Xu, Q.; Lekina, Y.; Ramesh, S.; Shen, Z. X.; Sum, T. C. Electronic states modulation by coherent optical phonons in 2D halide perovskites. Advanced Materials 2021, 33 (11), 2006233.

(6) Coropceanu, V.; Cornil, J.; da Silva Filho, D. A.; Olivier, Y.; Silbey, R.; Brédas, J.-L. Charge transport in organic semiconductors. Chemical reviews 2007, 107 (4), 926-952.